\documentstyle[aaspp4]{article}
\newcommand{\simgt}{\lower.5ex\hbox{$\; \buildrel > \over \sim \;$}}
\newcommand{\simlt}{\lower.5ex\hbox{$\; \buildrel < \over \sim \;$}}
\newcommand{\citet}[1] {\cite{#1}}
\newcommand{\citep}[1] {(\cite{#1})}
\newcommand{\bm}[1]{\mbox{\boldmath$#1$}}
\newcommand{\kaco}[1]{\left\langle{#1}\right\rangle}
\newcommand{\skaco}[1]{\langle{#1}\rangle}
\newcommand{\xipk}{\xi_{\rm pk-pk}}

\begin{document}
\title{Detectability of Gravitational Lensing Effect on the Two-point
Correlation Function of Hotspots in the CMB maps}

\author{\sc Masahiro Takada and Toshifumi Futamase}
\affil{%
Astronomical Institute, Graduate School of Science, 
Tohoku University, Sendai 980-8578, Japan}
\affil{%
takada@astr.tohoku.ac.jp; tof@astr.tohoku.ac.jp}

\begin{abstract}
We present quantitative investigations of the weak lensing effect on
the two-point correlation functions of local maxima (hotspots),
 $\xipk(\theta)$,  in the cosmic microwave background (CMB) maps.
The lensing effect depends on the projected mass fluctuations between
today and the redshift $z_{\rm rec}\approx1100$. 
If adopting the Gaussian assumption for the primordial temperature
 fluctuations field,
the peak statistics can provide an additional information about the
intrinsic distribution of hotspots that those pairs have
 some characteristic separation angles. The weak lensing
 then redistributes hotspots in the observed
CMB maps from the intrinsic distribution and consequently
imprints non-Gaussian signatures onto $\xipk(\theta)$.
Especially, since the intrinsic $\xipk(\theta)$ has a pronounced
depression feature around the angular scale of  $\theta\approx 70'$
for a flat universe, the weak lensing induces a large smoothing at the
scale. 
We show that the lensing signature therefore has an advantage
to effectively probe mass fluctuations with large wavelength modes
around $\lambda\approx 50 h^{-1}{\rm Mpc}$.  To reveal the 
detectability, we performed numerical experiments with
specifications of {\em MAP} and {\em Planck Surveyor} including
the instrumental effects of beam smoothing and detector noise.
It is then found that our method can successfully provide constraints
on amplitude of the mass fluctuations and cosmological parameters
in a flat universe with and without cosmological constant,
provided that we use maps with $65\%$ sky coverage expected from Planck.
\end{abstract}
%%%%%%%%%%%%%%%%%%%%%%%%%%%%%%%%%%%%%%%%%%%%%%%%%%%%%%%%%%%%%%%%%%%%%%%%
%%%%%%%%%%%%%%%%%%%%%%%%%%%%%%%%%%%%%%%%%%%%%%%%%%%%%%%%%%%%%%%%%%%%%%%%

 \section{Introduction}

  The temperature anisotropies in the comic microwave background (CMB)
contain detailed information about the underlying cosmological model
(\cite{HSS}).  
The recent high precision balloon-borne experiments,
{\sc Boomerang} (\cite{boom}; \cite{lange}) and MAXIMA
(\cite{Maxima}; \cite{balbi}), revealed that the measured
angular power spectrum $C_l$ are in good agreement with
that predicted by standard inflation paradigm.
On the other hand, the large-scale structure
of the universe imprints secondary effects on the primordial temperature
fluctuations. One of them is the weak lensing effect;
the CMB photons are randomly deflected by the gravitational field
due to the intervening
large-scale structure during the propagations from the last
scattering surface to a telescope.  The weak lensing can be a
powerful probe for mapping inhomogeneous distribution of dark matter
in the universe (\cite{Gunn}; see \cite{BSreview} for a review),
which is not directly attainable by any
other means. In fact, several groups (\cite{Ludovic}; \cite{Bacon};
\cite{wtk}; \cite{kaiser}) recently have reported significant detections of
the coherent distortion of faint galaxies images arising from
the gravitational 
lensing by the foreground large-scale structure, and showed that those
results can provide some constraints on the cosmological parameters. 
However, it would be still extremely interesting to be able to
measure the lensing effects on the CMB and the detection would
be very precise for constraining
the cosmological parameters because there is no ambiguity in
theoretical understanding of the unlensed CMB physics
and about the distance of the source plane. 

The inflationary scenarios also predict that the primordial temperature
fluctuations are Gaussian. In this case, 
the statistical properties of any unlensed CMB field can
be exactly predicted based on the Gaussian random theory developed by
Bardeen et al. (1986;  hereafter BBKS) and Bond \& Efstathiou (1987;
hereafter BE) for three- and two-dimensional cases, respectively.
However, the weak lensing then induces the non-Gaussian signatures
in the observed CMB maps. Based on these considerations, some
specific features on the lensed temperature fluctuations field have been 
revealed. Bernardeau (1998) investigated how the weak lensing alters the 
probability density function (PDF) of the ellipticities defined from the
local curvature matrix of the temperature fluctuations field. The unlensed
PDF indeed has specific statistical properties for
the two dimensional Gaussian field,
and then the gravitational distortion induces an excess of
elongated structures in the CMB maps in the similar way as
the distortions of distant galaxies. Although the method could be 
a powerful probe to measure the lensing signatures around the
characteristic curvature scale ($\sim 5'$) of the unlensed temperature field,
the instrumental
effect of a finite beam size is crucial for the detection because the beam
smearing effect again tends to circularize the deformed local
structures. Hence, Van Wearbeke, Bernardeau \& Benabed (1999) 
investigated how the weak lensing causes a coherent distribution of the
relative orientation between the CMB and distant galactic ellipticities,
and proposed that it can be a efficient tool because the
orientation of the CMB ellipticities is robust against the beam
smearing.
%Likewise it has been proposed that the geometrical quantities
%of the subset of CMB maps, which exceeds a certain threshold
%value of temperature fluctuations, could effectively measure
%the non-Gaussian signatures (\cite{JTF}). 

We (\cite{TKF}; hereafter TKF) recently investigated the
weak lensing effect on the two-point correlation function of local
maxima ({\em hotspots}), say $\xi_{\rm pk-pk}(\theta)$, 
in the two dimensional CMB maps. Since the distribution of hotspots
is a point process, the analysis focused on the secondary effect how
the weak lensing redistributes hotspots in the observed CMB maps from
the intrinsic distribution. The unlensed $\xipk(\theta)$ can be then 
accurately predicted by the Gaussian random theory once $C_l$ is given
(BE; \cite{HS} hereafter HS).
According to the acoustic peaks in the $C_l$, the pairs of hotspots have
some characteristic angular separations on the last scattering surface.
We then found that the weak lensing fairly smooths out
the oscillatory shape of $\xi_{\rm pk-pk}(\theta)$.
In particular, the most interesting result is that the
lensing contribution to $\xipk(\theta)$ at angular scales ($\approx
70'$) corresponding to the first Doppler peak of $C_l$ is
relatively large, and we thus expect that $\xipk(\theta)$ can be a
sensitive statistical tool to measure the projected mass fluctuations
at such larger scales than the other methods do.  
The crucial quantity of our method is the lensing fluctuations of relative
angular separation between two CMB photons, 
and the lensing signatures to $\xipk(\theta)$ at such large
scales is the consequence of large scale modes of lensing deflection
angles. The simulated maps indeed illustrate that each displacement of
peak positions in the lensed maps from the unlensed maps is relatively
large even though both global features of pattern of temperature fluctuations
nearly trace each other (\cite{Zald} and also see Figure \ref{fig:cmbmap}).
Furthermore, these considerations lead to the expectation that
our method is not particularly affected by the beam smearing effect. 

The purpose of this paper, therefore, is to quantitatively investigate
the weak lensing effect on $\xipk(\theta)$ and reveal in detail the physical
interpretations of the effect. For the practical purpose we
perform quantitative investigations of the problem
whether the future satellite missions,
{\em MAP}\footnote{{\tt http://map.gsfc.nasa.gov}}
and {\em Planck Surveyor}\footnote{{\tt
http://astro.estec.esa.nl/SA-general/Projects/Planck/}}, 
can measure the lensing signatures to $\xipk(\theta)$ for
constraining the cosmological parameters. 
This can be done by using numerical experiments of both
unlensed and lensed CMB maps including the instrumental effects of
finite beam size and detector noise.  

This paper is organized as
follows. In the next section, we introduce
the formalism to investigate the lensing effect on $\xipk(\theta)$.
In Section 4, the formalism is applied to some specific cosmological models,
and we then compute the signal-to-noise ratios of the lensing signatures 
to $\xipk(\theta)$ using the numerical experiments in Section
5. Finally, in Section 6 we present discussions and conclusions. 

\section{Weak Lensing Effect on Two-point Correlations of Hotspots in
 the CMB: Formalism}

A bundle of CMB photons is randomly deflected by the inhomogeneous
matter distributions of the intervening large-scale structure as 
it propagates from the last scattering surface to a telescope. 
The two CMB bundles observed 
with a certain angular separation $\theta$ thus have a different angular
separation when emitted from the last scattering surface.
The ensemble averages of the second moment of the relative separation
fluctuations and the
following characteristic function can be then easily
calculated by using the power spectrum approach developed by Seljak (1996); 
\begin{eqnarray}
&&\sigma_{\rm GL}^2(\theta)\equiv2^{-1}
 \kaco{(\delta\bm{\theta}_1-\delta\bm{\theta}_2)^2}
 _{|\bm{\theta}_1 -\bm{\theta}_2|=\theta}
 =\sigma_{\rm GL,0}^2(\theta)+\sigma^2_{\rm GL,2}(\theta),
 \nonumber \\
&&\kaco{\exp[i\bm{l}\cdot(\delta\bm{\theta}_1-\delta\bm{\theta}_2)]}
_{|\bm{\theta}_1 -\bm{\theta}_2|=\theta}\simeq
1-\frac{l^2}{2}\left[\sigma^2_{\rm GL,0}(\theta)
+\cos(2\varphi_l)\sigma^2_{\rm GL,2}(\theta)\right],
\label{eqn:avergl}
\end{eqnarray}
where $\delta \bm{\theta}_1(\equiv \delta\bm{\theta}(\bm{\theta}_1))$
and $\delta \bm{\theta}_2(\equiv \delta\bm{\theta}(\bm{\theta}_2))$
are the angular
excursions of the two bundles, and $\kaco{\ \ }_\theta$ observationally
means the average performed over all pairs with a fixed observed angular
separation $\theta$. $\sigma_{\rm GL,0}(\theta)$ and $\sigma_{\rm
GL,2}(\theta)$ represent isotropic and anisotropic contributions to the
lensing dispersion, respectively. Although the anisotropic one is
ignored in TKF for simplicity, we also take into account the contribution
in this paper. It is convenient to express those dispersions in terms of 
the logarithmic angular power spectrum of the deflection angle, $P_{\rm
GL}(l)$, as
\begin{eqnarray}
&&\sigma_{\rm GL,0}^2(\theta)=\frac{1}{2\pi}\int\!\!\frac{dl}{l}
 P_{\rm GL}(l)[1-J_0(l\theta)], \nonumber \\
&&\sigma_{\rm GL,2}^2(\theta)=\frac{1}{2\pi}\int\!\!\frac{dl}{l}
 P_{\rm GL}(l)J_2(l\theta),
\label{eqn:gldisp}
\end{eqnarray}
with 
\begin{equation}
P_{\rm GL}(l)=9H^4_0\Omega_{\rm m0}^2\int^{\chi_{\rm rec}}_0\!d\chi
 a^{-2}(\tau)W^2(\chi,\chi_{\rm rec})
 P_\delta\left(k=\frac{l}{r(\chi)},\chi\right).
 \label{eqn:lensps}
\end{equation}
The statistical properties of the lensing effects are
thus entirely determined by $P_{\rm GL}(l)$, because the lensing field
is expected to be also Gaussian at relevant angular
scales ($\theta\simgt 10'$). $\tau$ is a conformal time, $\chi\equiv \tau_0-\tau$,
$J_n(x)$ is the Bessel function of order $n$, and 
the subscript $0$ and ``rec'' denote values at present and a
recombination time, respectively. $P_\delta(k,\tau)$ is the power
spectrum of matter fluctuations field, $H_0=100h$ km s$^{-1}$ Mpc$^{-1}$
and $\Omega_{\rm m0}$ denote the present Hubble parameter
and the present energy density of matter, respectively.
$r(\chi)$ is the corresponding comoving angular diameter distance,
defined as $K^{-1/2}\sin K^{1/2}\chi$, $\chi$,
$(-K)^{-1/2}\sinh (-K)^{1/2}\chi$ for $K>0, K=0, K<0$, respectively,
where the curvature parameter $K$ is represented as
$K=(\Omega_{\rm m0}+\Omega_{\lambda 0}-1)H_0^2$ and $\Omega_{\lambda 0}$ 
is the present vacuum energy density.
The projection operator $W(\chi,\chi_{\rm rec})$ on the celestial sphere
is given by $W(\chi,\chi_{\rm rec})=r(\chi_{\rm rec}-\chi)/r(\chi_{\rm rec}$).
In the derivation of equation (\ref{eqn:avergl}), we have employed two
approximations. First is the flat-sky approximation
where the two dimension Fourier transformation is used neglecting the
curvature of the celestial sphere. This is based on the
consideration that the lensing are important only on small angular
scales. Second is the Born approximation that the integral can be
evaluated along the unperturbed null-geodesics of CMB photon.
Hu (2000) recently investigated the correction to the flat sky approach
by directly evaluating the lensing
effect in harmonic space. This correction to our method, however, is
small as will be discussed later. The Born approximation is valid as
long as the $\sigma_{{\rm GL}}(\theta)/\theta\ll 1$ is satisfied, and  
we numerically  confirmed this on the relevant 
angular scales for all cosmological models considered in this paper.  
Importantly, magnitude of the lensing dispersion (\ref{eqn:avergl}) is
particularly sensitive to $\Omega_{\rm m0}$ and the normalization of
matter power spectrum, which is conventionally expressed in terms of
the rms mass fluctuations of a sphere of $8h^{-1}$Mpc, i.e.,
$\sigma_8$. 

The hotspots are local maxima in the two dimensional CMB sky map,
and hence the distribution obeys a point process.
Once the angular power spectrum
of the temperature fluctuations field $C_l$ is given,
the Gaussian assumption allows us to
exactly predict statistical properties of the intrinsic distribution
of hotspots following the methods developed by BBKS and BE.
Since $C_l$ has oscillatory features such as series of Doppler peaks,
the pairs of hotspots are distributed with some {\em characteristic}
separation angles on the last scattering surface as shown by HS and
TKF. This result leads to the following expectation. Let us consider
all pairs of hotspots separated with the certain characteristic angular scale.
Although all those pairs should be observed with the characteristic
scale in the absence of the lensing, they are actually observed with
various different separations in random lines of sight
because of the weak lensing.
The probability distribution of observed separation angles
then has the lensing dispersion (\ref{eqn:avergl}) at
the characteristic scale. 
Since the effect can be measured in only a statistical sense,
in this paper we focus on investigations of the lensing effect on the
{\em two-point} correlation function of hotspots.
Note that our method do not consider spurious hotspots created by
the lensing, but it is a good approximation
because it has been shown that an additional
power of the anisotropies generated by the weak lensing is
very small and important only at small scales ($\theta\simlt 1'$)
(\cite{Metcalf97}; \cite{SZ99}; \cite{Zald}).
These features are indeed illustrated by the
numerically simulated CMB maps (see Figure \ref{fig:cmbmap}) 

To calculate the weak lensing effect on the two-point correlation
function of hotspots in the CMB maps,
we first define the number density fluctuations field
of hotspots as
\begin{equation}
\delta n_{\rm pk}(\bm{\theta})=\frac{n_{\rm pk}(\bm{\theta})
 -\bar{n}_{\rm pk}}{\bar{n}_{\rm pk}}, 
\end{equation}
where $n_{\rm pk}(\bm{\theta})$ and $\bar{n}_{\rm pk}$ are
the number density field and the mean number density of hotspots above
a certain threshold $\nu$, respectively. The threshold is conventionally 
expressed in units of the rms temperature
fluctuations as $\nu=\Delta_{\rm pk}/\sigma_0$, where
$\Delta_{\rm pk}$ is the value of temperature fluctuation at the
peak position defined by $\Delta(\bm{\theta}_{\rm pk})\equiv\delta
T(\bm{\theta}_{\rm pk})/T_{\rm CMB}$ and  the dispersion is defined by
$\sigma_0^2\equiv\skaco{\Delta^2(\bm{\theta})}$. Similarly,
the other spectral parameters are defined by
$\sigma_1^2\equiv\skaco{(\nabla\Delta)^2}$ and $\sigma_2^2\equiv
\skaco{(\nabla^2\Delta)^2}$.  These parameters can be expressed in
terms of $C_l$ in the context of the small angular approximation (BE) as
$\sigma_n^2\equiv\int(ldl/(2\pi))C_l l^{2n}$. 
The analytical expression of $\bar{n}_{\rm pk}(>\nu)$
has been derived by BE and is in detail presented in appendix
\ref{app:num}. Because of the mapping effect due to weak lensing, the
lensed (observed) fluctuations field,
$\delta n^{\rm GL}_{\rm pk}(\bm{\theta})$,  at a certain angular position
$\bm{\theta}$ is the intrinsic field at another position
$\bm{\theta}+\delta\bm{\theta}$
on the last scattering surface, where $\delta
\bm{\theta}$ is the deflection angle. Thus $\delta
n^{\rm GL}_{\rm pk}(\bm{\theta})$ can be expressed as 
\begin{equation}
\delta n^{\rm GL}_{\rm pk}(\bm{\theta})=\delta n_{\rm pk}(\bm{\theta}
 +\delta\bm{\theta})=\int\!\!\frac{d^2\bm{l}}{(2\pi)^2}\delta n_{\bm{l}}e^{i\bm{l}
 \cdot(\bm{\theta}+\delta\bm{\theta})},
\end{equation}
where $\delta n_{\bm{l}}$ is the Fourier component of unlensed field
$\delta n(\bm{\theta})$. 
Therefore, since the lensing deflection angle induced by the large-scale 
structure and the CMB field on the last scattering surface
are statistically independent,
the lensed (observed) two-point correlation function of hotspots,
$\xi^{\rm GL}_{\rm pk-pk}(\theta)$, can be calculated with help of
equation (\ref{eqn:avergl}) as
\begin{eqnarray}
\xipk^{\rm GL}(\theta)&=&\kaco{\delta n^{\rm GL}(\bm{\theta}_1)
 \delta n^{\rm GL}(\bm{\theta}_2)}_{|\bm{\theta}_1-\bm{\theta}_2|=\theta}
\nonumber \\
 &=& \int\!\!\frac{d^2\bm{l}}{(2\pi)^2}\int\!\!\frac{d^2\bm{l}'}{(2\pi)^2}
  e^{i(\bm{l}\cdot\bm{\theta}_1-\bm{l}'\cdot\bm{\theta}_2)}
  \kaco{\delta n_{\bm{l}}\delta n_{\bm{l}'}}\kaco{e^{i(\bm{l}\cdot \delta\bm{\theta}_1
  -\bm{l}'\cdot\delta\bm{\theta}_2)}}\nonumber \\
 &=& \int^\infty_0 \!\frac{ldl}{2\pi}P_{\rm pk-pk}(l)\left[
 \left(1-\frac{l^2}{2}\sigma^2_{\rm GL,0}(\theta)\right)J_0(l\theta)
 +\frac{l^2}{2}\sigma^2_{\rm GL,2}(\theta)J_2(l\theta)\right], \label{eqn:xigl0}
\end{eqnarray}
where we have used the 
following Gaussian random property of $\delta n_{\bm{l}}$;
\begin{equation}
\kaco{\delta n_{\bm{l}}\delta n_{\bm{l}'}}=(2\pi)^2P_{\rm pk-pk}(l)
 \delta^2(\bm{l}-\bm{l}').
\end{equation}
$P_{\rm pk-pk}(l)$ is the angular power spectrum of unlensed two-point
correlation function of hotspots, $\xipk(\theta)$.
The derivation of $\xipk(\theta)$ is presented in detail in the appendix
\ref{app:unlensxi}.
The relation between $\xipk(\theta)$ and $P_{\rm pk-pk}(l)$ can be expressed as
\begin{equation}
P_{\rm pk-pk}(l)=2\pi\int_0^\pi\! \theta d\theta \xipk(\theta)J_0(l\theta). 
\label{eqn:powerxi}
\end{equation}
By using this equation, equation (\ref{eqn:xigl0}) can be therefore rewritten as
\begin{equation}
\xipk^{\rm GL}(\theta)=\int\!\!d\theta'\theta'\int\!\!ldl\xi_{\rm pk-pk}(\theta')
 J_{0}(l\theta')\left[
 \left(1-\frac{l^2}{2}\sigma^2_{\rm GL,0}(\theta)\right)J_0(l\theta)
 +\frac{l^2}{2}\sigma^2_{\rm GL,2}(\theta)J_2(l\theta)\right].
\label{eqn:lensxi}
\end{equation}
This is the equation which we use for theoretical predictions of the 
lensing effect on $\xipk(\theta)$. 
If we ignore the anisotropic lensing dispersion $\sigma_{\rm GL,2}$,
the expression (\ref{eqn:lensxi}) can be further simplified (see
appendix \ref{app:xianaly}).
Importantly, this equation indicates that the lensing contribution to
$\xipk(\theta)$ at a certain scale $\theta$ arises only from the lensing
dispersion $\sigma_{\rm GL}(\theta)$ at the same scale. Hence,  
detections of scale dependences of the lensing signatures to $\xipk(\theta)$
allow us to reconstruct the lensing dispersion at the respective scales,
more interestingly, to reconstruct the projected matter power spectrum. 

\section{Cosmological models}

To make some quantitative predictions, one needs to specify
cosmological models. For this reason, we adopt
following adiabatic cold dark matter models with
$\Omega_{\rm m0}=1$,
$\Omega_{\rm \lambda 0}=0$, $h=0.5$ (hereafter SCDM) and $\Omega_{\rm m0}=0.3$,
$\Omega_{\rm \lambda 0}=0.7$, $h=0.7$ (hereafter LCDM), respectively.
These models are motivated by the fact that
the recent high precision measurements of $C_l$
supported a flat universe under the adiabatic condition
as suggested by standard inflationary scenarios (\cite{boom}; \cite{Maxima}). 
The baryon density is chosen to satisfy $\Omega_{\rm b0}h^2=0.019$, which
is consistent with values obtained from the measurements of the primeval 
deuterium abundance (\cite{tytler}). As for the
matter power spectrum, we employed the Harrison-Zel'dovich spectrum and
the BBKS transfer function with the shape parameter from Sugiyama
(1995). The free parameter in each model is only the normalization of the
present-day matter power spectrum, i.e., $\sigma_8$.
The nonlinear evolution of the power spectrum can be modeled using the fitting
formula given by Peacock \& Dodds (1996).
To compute the angular power spectrum $C_l$,
we used helpful CMBFAST code developed by Seljak \& 
Zaldarriaga (1996). 

\section{Theoretical Results}

\subsection{Dependence of lensing dispersion on cosmological parameters}

%%%%%%%%%%%%%%%%%%%%%%%%%%%%% Figure 1 %%%%%%%%%%%%%%%%%%%%%%%%%%%%%%%%%
\begin{figure}[t]
 \begin{center}
     \leavevmode\epsfxsize=10cm \epsfbox{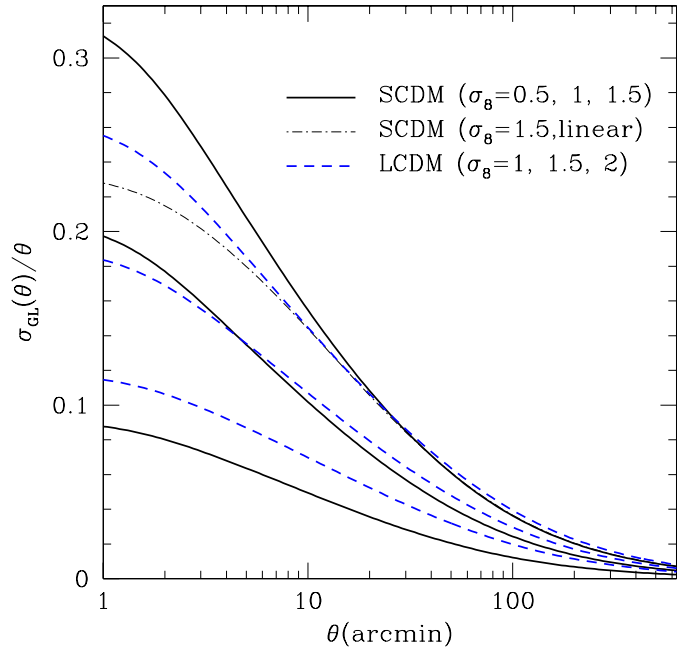}
\caption{
  The dependence of $\sigma_{\rm GL}(\theta)/\theta$ on $\theta$ is shown for
  two cosmological models with various normalizations of CDM power
  spectrum and $\Omega_{\rm b0}h^2=0.019$.
  The solid lines show results of $\sigma_8=0.5, 1.0$ and $1.5$
  in SCDM ($\Omega_{\rm m0}=1$ and  $h=0.5$) from bottom to top
  while the dashed lines are $\sigma_8=1.0, 1.5$ and $2.0$ in LCDM
  ($\Omega_{\rm m0}=0.3$, $\Omega_{\rm \lambda 0}=0.7$), respectively, 
  where the nonlinear effect on the matter power spectrum
  is computed using the fitting formula by Peacock \& Dodds
  (1996). The thin dot-dashed line is for the
  linear power spectrum in SCDM with $\sigma_8=1.5$.
\label{fig:dispgl}}
 \end{center}
\end{figure}

In Figure \ref{fig:dispgl} we plot the relative lensing
dispersion $\sigma_{\rm GL}(\theta)/\theta$ as a function of the
separation angle $\theta$
for different sets of $\sigma_8$ in SCDM and LCDM
models, which are computed using equations (\ref{eqn:avergl}).
The dependence of $\sigma_8$ in each model is demonstrated by choices of
$\sigma_8=0.5, 1$, and $1.5$ (solid lines) in SCDM,
and $\sigma_8=1, 1.5$ and $2.0$ (dashed lines) in LCDM from bottom to top,
respectively. The normalizations from the {\em COBE} 4-year measurements
(e.g. \cite{Bunn}) and the {\em X}-ray cluster abundance (\cite{Eke};
\cite{KS}) roughly correspond to $\sigma_8=1.2$ and $0.5$ for SCDM,
and $\sigma_8\simeq 1-1.5$ for LCDM, respectively. Furthermore, the
recent several measurements of cosmic shear (e.g. \cite{Ludovic})
have suggested $\sigma_8=1.5\pm 0.5$ for the current favored LCDM models,
while uncertainties involved in redshift distributions of distant
galaxies, the cosmic variance of the variance of the shear and
the systematic errors of the signals still remain unresolved. 
Figure \ref{fig:dispgl} clearly shows that the magnitude
of $\sigma_{\rm GL}(\theta)$ at a certain angle $\theta$
has a strong dependence
on the amplitude of $\sigma_8$ in each cosmological model.
Since the logarithmic angular power spectrum
(\ref{eqn:lensps}) of $\sigma_{\rm GL}$
has contributions from the proportional factor $\Omega^2_{\rm m0}$ and
the distance and the growth factors, the combination 
yields a dependence of $\Omega_{\rm m0}$ on $\sigma_{\rm GL}(\theta)$.  
Although the Hubble parameter $h$ affects $\sigma_{\rm GL}$ mainly
through the shape parameter $\Gamma \approx \Omega_{\rm m0}h$
in the matter power spectrum, 
the dependence is weaker. The thin dot-dashed line shows a result of using 
the linear matter power spectrum for $\sigma_8=1.5$ in SCDM,
and it reveals that the effect of the nonlinear evolution on
$\sigma_{\rm GL}$
is not important at $\theta\simgt 10'$, where the two-point correlations
function of hotspots has most power of correlations.
It is therefore expected that 
the measurements of $\sigma_{\rm GL}$ can generally provide constraints on
 $\sigma_8-\Omega_{\rm m0}$ plane. To break the degeneracy
 between $\Omega_{\rm m0}$ and $\sigma_8$, we have to
 measure the scale dependence of $\sigma_{\rm GL}(\theta)$ with
 respect to $\theta$.
%comparisons of $\sigma_{\rm GL}$ measurements at different several scales.

%%%%%%%%%%%%%%%%%%%%%%%%%%% Figure 2%%%%%%%%%%%%%%%%%%%%%%%%%%%%%%%%%%%
\begin{figure}[t]
 \begin{center}
     \leavevmode\epsfxsize=10cm \epsfbox{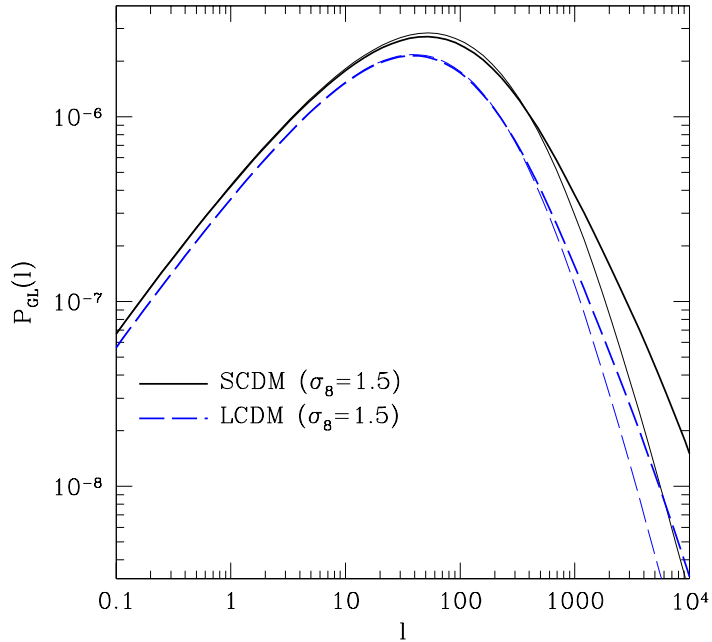}
\caption{
The logarithmic angular power spectrum
of the lensing dispersion as a function of $l$ for SCDM (solid) and LCDM 
(dashed) models with $\sigma_8=1.5$ as in Figure \ref{fig:dispgl}.
The power spectrum is  computed using equation (\ref{eqn:lensps}). 
The thin and thick lines in each model show the corresponding 
results of using linear and nonlinear three dimensional
power spectrum of matter fluctuations, respectively.
\label{fig:2dps}
}
\end{center}
\end{figure}

As shown by equations (\ref{eqn:gldisp}) and (\ref{eqn:lensps}),
the lensing dispersion arises from the projected
matter power spectrum $P_{\delta}(k)$ weighted
with $a^{-2}W^2(\chi,\chi_{\rm rec})$.
Figure \ref{fig:2dps} shows the logarithmic angular power
spectrum $P_{\rm GL}(l)$ defined by equation (\ref{eqn:lensps})
as a function of $l$. The thin and thick curves are
results of using the linear and nonlinear 
matter power spectra, respectively, and the figure
demonstrates that the nonlinear effect is important on $l\simgt 1000$, which
corresponds to angular scales of
$\theta\simlt 20'$ from the relation of $l\approx
2\pi/\theta$.
The shape of $P_{\rm GL}$ peaks around $l\approx 100$ and 
thus the contributions to the deflection angle of {\em each} CMB photon
come mainly  from modes with such large $l$ (see Figure \ref{fig:cmbmap}).
%This feature is actually illustrated by
%Figure \ref{fig:cmbmap}.
%For the lensing effect on the two-point
%correlation function of hotspots
%As shown by equation (\ref{eqn:lensxi}), the lensing flcutatoins of
%relative separation angle then produce contributions of the
%lensing effect on the two-point correlation function of hotspots.  
The essential quantity of lensing effect on $\xipk$ is lensing
fluctuations of relative separation angle,
and the term including Bessel function in $\sigma_{\rm GL,0}(\theta)$
of equation (\ref{eqn:gldisp})
indicates that the lensing fluctuations for two CMB bundles
separated with a certain angle $\theta$ arises dominantly from the
integrations of $P_{\rm GL}(l)$ over $l>2\pi/\theta$ in $l-$space.
This physically means that the gravitational lensing effect on the
relative angular separation is caused mainly by the projected matter
fluctuations lied {\em between} the two bundles.
Accordingly the two bundles with smaller separation are
more strongly affected by the smaller scale structures of the universe.
Furthermore, from these interpretations we can conclude
that corrections to the flat sky approximation proposed by Hu (2000)
do not affect our results at $l\simgt 100$ because the corrections are
important only at $l<10$.  
%It will be here worth to remark differences between our method
%and other methods. For investigations of the gravitational deformation
%effect on the CMB maps and images of distant galaxies, the convergence field, more 
%strict speaking the second derivatives field of the gravitational
%potential, plays a crucial role. The convergence field is then given by
%the radial integral of three dimensional density fluctuations
%field weighted with $W r(\chi)/a$, and has a logarithmic angular
%power spectrum proportional to $l^{2}P_{\rm GL}(l)$ as shown in equation 
%(\ref{eqn:pskappa}). 
%The deformation effect is therefore more sensitive
%to larger-$l$ mode, namely, to the matter fluctuations with smaller
%wavelengths.  The curve of $l^{2}P_{\rm GL}(l)$
%in $l$-space is indeed a monotonous increasing function
%(\cite{JS}; \cite{ZS99}). Our method thus has an advantage to probe more 
%effectively the matter power spectrum at relatively large angular
%scales such as $\theta\approx 70'$ by using
%the lensing fluctuations of relative angular separations.

%%%%%%%%%%%%%%%%%%%%%%%%%%% Figure 3 %%%%%%%%%%%%%%%%%%%%%%%%%%%%%%%%%%%
\begin{figure}[t]
 \begin{center}
     \leavevmode\epsfxsize=14cm \epsfbox{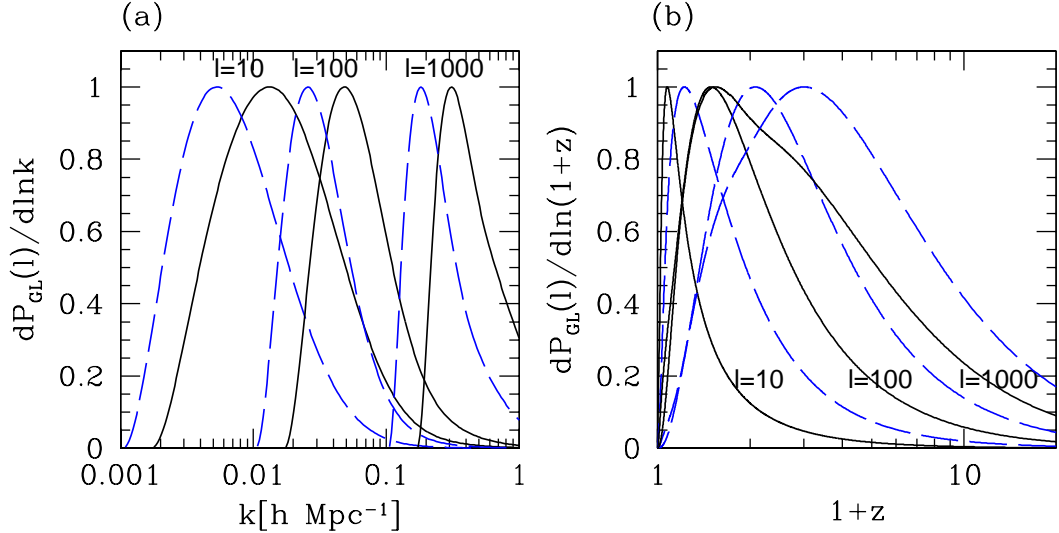}
\caption{
 (a) The logarithmic contribution to the
  2D power spectrum $P_{\rm GL}(l)$ defined by equation
  (\ref{eqn:lensps}) as a function of $k$ for $l=10$,
  $100$, $1000$, where the normalization is arbitrarily scaled.
  The models are same as those in Figure \ref{fig:2dps}. 
  (b) The logarithmic contribution to $P_{\rm GL}(l)$ for
  each $l$ mode as a function of $1+z$. 
\label{fig:psdkdz}
}
\end{center}
\end{figure}
Since the more fundamental quantity 
is the three dimensional density fluctuations characterized by the
power spectrum $P_\delta(k)$, we have to see the relation between the two power
spectra, namely $P_\delta(k)$ and $P_{\rm GL}(l)$.
Figure \ref{fig:psdkdz}(a) shows the logarithmic
contribution to $P_{\rm GL}(l)$ for a given $l$-mode as a function of
three dimensional wavenumber $k$ in the same models as in Figure
\ref{fig:2dps}, where normalizations of those curves are arbitrarily scaled.
These functions have relatively broad shape and
peaks at $\lambda=k/2\pi\approx 630 h^{-1}{\rm Mpc}$ for $l=10$, at
$k\approx 125h^{-1}{\rm Mpc}$ for $l=100$, and $k\approx 21h^{-1}{\rm
Mpc}$ for $l=1000$ in the considered SCDM model, respectively.
These relations between $l$ and $k$
depend on the shape of matter power spectrum. On the other hand,
since in the
LCDM model the comoving angular diameter distance at a certain
redshift is larger than the corresponding distance
in SCDM, the peak-wavelengths are smaller
than those in the SCDM cases for each $l$ mode.
The figure also demonstrates that the higher $l$ mode is affected
more strongly by the matter fluctuations with smaller wavelengths. 

The next question is the redshift distribution of
the contribution to a given $P_{\rm GL}(l)$. 
Figure \ref{fig:psdkdz}(b) shows the logarithmic contribution
to $P_{\rm GL}(l)$ for each $l$ mode as a function of $(1+z)$.
Since the window function $W/a$ in $P_{\rm GL}(l)$ is a monotonous
decreasing function with respect to $z$ and has no characteristic redshift,
the question which redshift structures give the dominant
contribution depends on the shape of matter power spectrum.
The figure clearly demonstrates that each curve peaks at a certain redshift;
for low $l$ mode the contribution is dominated by the low-$z$ structures
and, on the other hand, high $l$ mode has wide  range contributions
in the redshift space. These results can be explained as follows. 
As shown in Figure \ref{fig:psdkdz}(a), there is the peak scale $k_{\rm
max}$ of three dimensional mass fluctuations
that provides most dominant contribution
to $P_{\rm GL}(l)$ for each $l$ mode. 
Because of the projection effect on the celestial sphere,
the contribution from the $k_{\rm max}$ mode
comes from structures at a specific redshift $z_{\rm max}$ which satisfies
the relation of $k_{\rm max} \approx l/r(z_{\rm max})$. 
As a result, for lower $l$ mode the contribution will be
dominated by lower $z$ structures. On the other hand, for $l=1000$ mode
the contribution peaks at $z\approx 0.5$ with a long tail expanding to 
higher $z$, where the tail is caused by the fact that the window
function $W/a$ has larger power at lower $z$. In fact,
Figure \ref{fig:psdkdz}(b) confirms these interpretations.
We could therefore directly probe dark matter clustering
at low and high  redshifts for low and high $l$ modes, respectively,
in principle by using the weak lensing signatures.

\subsection{Unlensed two-point correlation function of hotspots
  with the instrumental effects}

%%%%%%%%%%%%%%%%%%%%%%%%%% Figure 4 %%%%%%%%%%%%%%%%%%%%%%%%%%%%%%%%%%%%
\begin{figure}[t]
 \begin{center}
     \leavevmode\epsfxsize=10cm \epsfbox{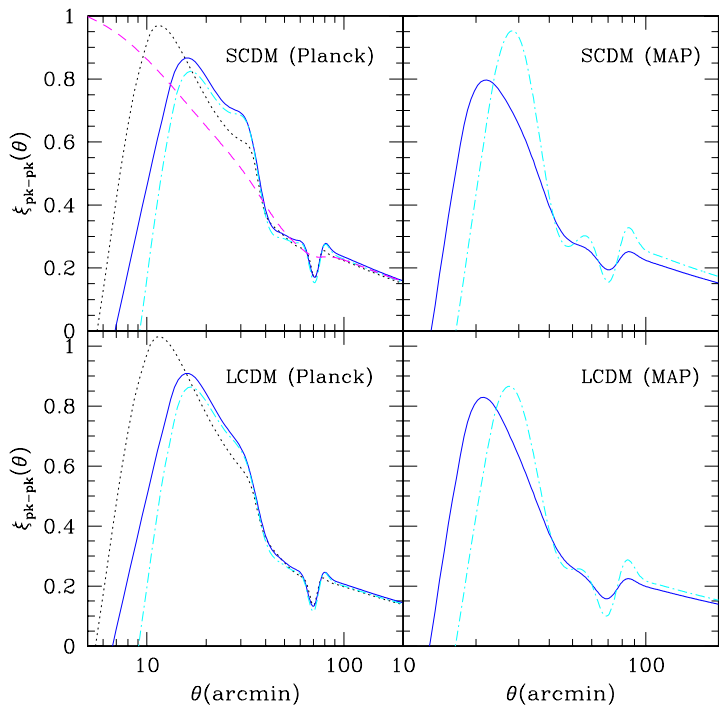}
\caption{
The unlensed two-point correlation function of hotspots of height
  above threshold $\nu=1$, $\xipk(\theta)$,
as a function of the separation angle $\theta$ for SCDM (upper panels) and 
LCDM (bottom panels). The instrumental effects of finite beam size and
detector noise for the expected Planck and MAP surveys are taken into account.
The solid and dot-dashed lines shows the cases including both those
effects and only the beam smearing effect, respectively. For
comparison, the ideal case of $\xipk(\theta)$ without those effects is shown by
the dotted lines in the upper left and bottom left figures.
The dashed line in the upper left figure shows the conventionally used
  two-point correlation function $C(\theta)$ normalized to agree with
 the value of $\xipk(\theta)$ with the solid line at
  $\theta=200'$. 
\label{fig:xipk} 
}
\end{center}
\end{figure}
The unlensed two-point correlation functions of local maxima ({\em
hotspots}),  say $\xi_{\rm pk-pk}(\theta)$, in the CMB maps can be
accurately predicted based on the Gaussian random theory. 
The derivation is presented in the appendix \ref{app:unlensxi} 
in detail. To show the shape of $\xipk(\theta)$,
we performed a six dimensional numerical integration 
using equation (\ref{eqn:unlensxi}) because the two of the eight dimensional
integration in equation (\ref{eqn:unlensxi}) can be analytically done. 
For the practical purpose, we also take into account
instrumental effects of finite beam size and detector noise.
The beam smearing effect on the temperature fluctuations field
can be modeled by the Gaussian beam approximation characterized by
a filter function $F_l=\exp[-l^2
\theta_s^2/2]$ in $l$-space, where the smoothing angle $\theta_s$
is expressed in
terms of the full-width at half-maximum angle $\theta_{\rm fwhm}$ of a
telescope as $\theta_s=\theta_{\rm fwhm}/\sqrt{8\ln 2}$.
The noise level of detectors is conventionally expressed in terms of
the temperature fluctuations per a pixel on a side of FWHM extent as
 $\Delta_{\rm noise}=\sigma_{\rm sens}$.
If we assume that the primordial temperature and the noise fields
are uncorrelated, by modifying the angular power spectrum
$C_l$ to $\tilde{C}_l=(C_l+\sigma_{\rm sens}^2\theta^2_{\rm
fwhm})\exp[-l^2\theta_s^2]$, 
these instrumental effects can be approximately included 
into the theoretical predictions (HS), 
because in the Gaussian random theory $C_l$ contains complete
information about statistical properties of any intrinsic CMB field.
The numerical experiments indeed show
that this treatment works well.

%%%%%%%%%%%%%%%%%%%%%%%%%%%% Figure 5 %%%%%%%%%%%%%%%%%%%%%%%%%%%%%%%%%%%%
\begin{figure}[t]
 \begin{center}
     \leavevmode\epsfxsize=14cm \epsfbox{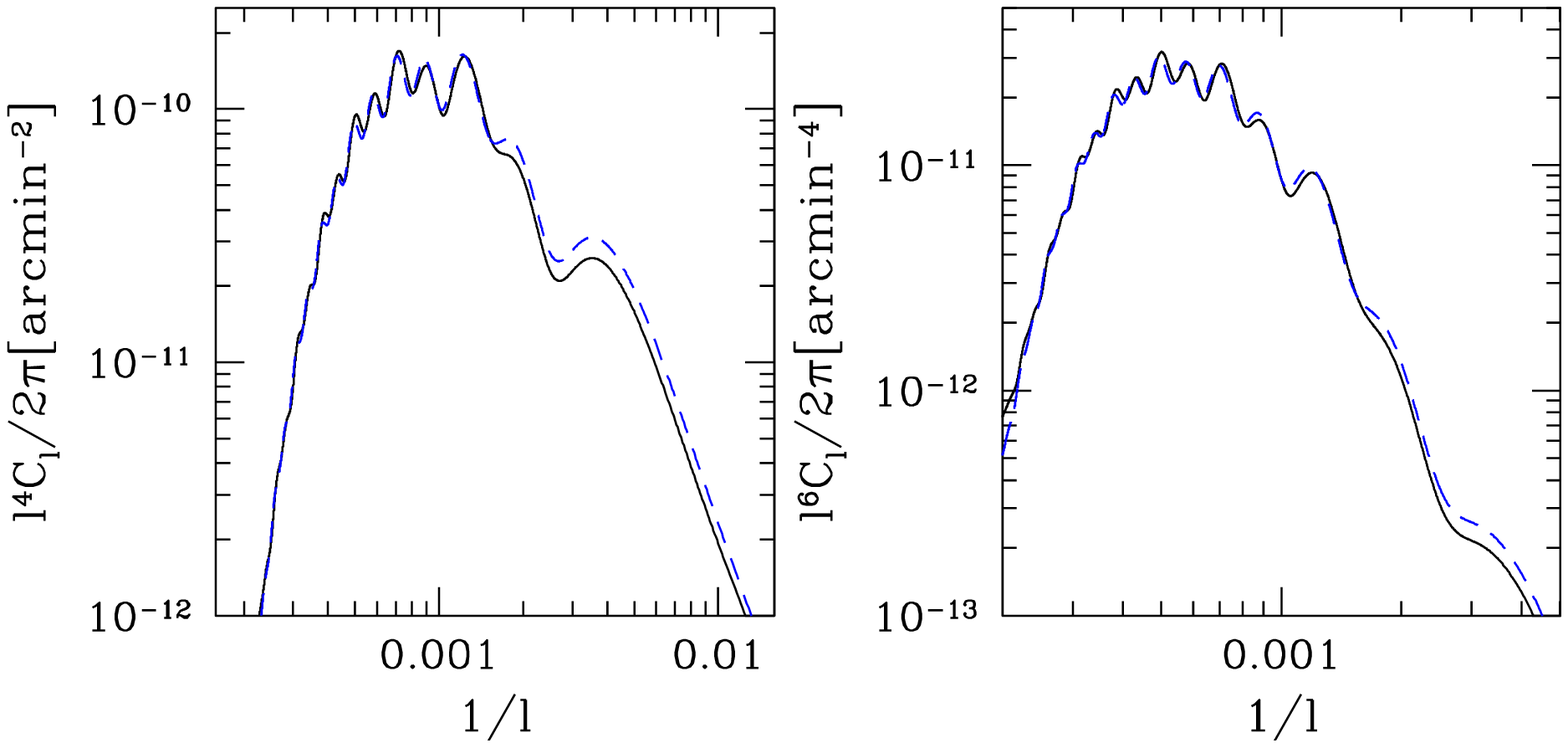}
\caption{
 The logarithmic power spectra of the gradient (left) and
  second-derivative (right) fields of the CMB anisotropies field as a function
  of $1/l (\propto \theta)$. The solid and dashed lines show the results 
   of SCDM and LCDM models, respectively. 
  This figure clearly explains that the oscillations of $l^2C_l$ is strongly
  enhanced. 
\label{fig:Clln}}
\end{center}
\end{figure}
Figure \ref{fig:xipk} shows the unlensed $\xipk(\theta)$ with and
without the experimental effects as a function of separation angle
$\theta$ in the SCDM and LCDM models, where we considered the
two hotspots of height above the threshold $\nu=1(\Delta_{\rm
pk}=\sigma_0)$.
%Note that, altough we adopt the treshold of $\nu=1$ for
%simplicity throughout this paper, the shape of $\xipk$ with an arbitrary
%threshold can be predicted accurately. 
As for the instrumental effects, we have employed the 
specifications of Planck $217{\rm GHz}$ channel and MAP $90{\rm GHz}$ channel;
the beam size and noise level are assumed to be 
$\theta_{\rm fwhm}=5.5'$ and $\sigma_{\rm sens}=4.3\times 10^{-6}$
for Planck while $\theta_{\rm fwhm}=12.6'$ and $\sigma_{\rm sens}=2.5\times
10^{-5}$ for MAP. The solid lines in each panel show
$\xipk(\theta)$ including both effects of the beam size and noise,
and the dotted lines in the upper left and bottom left panels
show an ideal case without those effects. The ideal cases demonstrate
that the intrinsic $\xipk(\theta)$ 
has a prominent peak at $\theta\approx 10'$ and
the damping tail at $\theta<10'$. These features physically mean that the 
primary temperature field has the characteristic curvature scale of
the order of $10'$, which can be estimated by $\theta_\ast\equiv
\sqrt{2}\sigma_1/\sigma_2 \sim 5'$ in Appendix \ref{app:num}, and
has smooth structures at scales of $\theta<\theta_\ast$ as actually shown 
by the simulated maps.  The beam smearing on the intrinsic
$\xipk(\theta)$ then appears as a cutoff at scales below the beam size
although the effect moderately changes the global shape of $\xipk(\theta)$
in the MAP case (top right and bottom right panels). 
This is because the beam smearing causes an incorporation of intrinsic
hotspots contained within one beam. 
To clarify the noise effect explicitly, we also show
$\xipk(\theta)$ only with beam smearing effect (dot-dashed lines).
The curves explain that spurious hotspots due to the detector noises
generate an extra power of correlations on $\xipk(\theta)$. 
In particular, the predicted shape of $\xipk(\theta)$ for
MAP is largely affected by the noise effect up to $\theta=80'$ while
the Planck cases have slight changes only at
$\theta\simlt20'$. By using equation (\ref{eqn:meannum}), we can predict
values of mean number density of hotspots above $\nu=1$ for those cases. 
The values with both the beam smearing and noise effects and only with beam
effect in SCDM model are $\bar{n}_{\rm pk}=8.17\times 10^3$ and
$7.74\times 10^3[{\rm rad}^{-2}]$ for Planck (upper left),
respectively, while $\bar{n}_{\rm pk}=5.31\times10^3$ and
$3.01\times10^3[{\rm rad}^{-2}]$ for MAP (upper right), respectively. 
Similarly, in LCDM model $\bar{n}_{\rm pk}=7.73\times 10^3$
and $7.33\times 10^3[{\rm rad}^{-2}]$ for Planck (bottom left) and
$\bar{n}_{\rm pk}=5.06\times 10^2$ and $2.88\times 10^3 [{\rm
rad}^{-2}]$ for MAP (bottom right), respectively.
These values actually coincide with results of the number counts of
 hotspots in the simulated CMB maps within the Poisson error. 
Moreover, to reveal the differences of shapes between $\xipk(\theta)$ and the
conventionally used two-point correlation function of the temperature
fluctuations field itself defined by $C(\theta)\equiv
\kaco{\Delta(\bm{\theta}_1)
\Delta(\bm{\theta}_2)}_{|\bm{\theta}_1-\bm{\theta}_2|=\theta}$, 
the dashed line in upper left panel shows $C(\theta)$
normalized to agree with the value of solid line at $\theta=200'$.
It is clear that $\xipk(\theta)$ 
has much more oscillatory features than $C(\theta)$ does (HS; TKF).
%The reason why $\xipk(\theta)$ has such oscillatory shapes 
%can be then explained as follows. 
This reason is as follows. 
In the peak statistics we need statistical properties of
the gradient and second derivative fields of the temperature
fluctuations field in order to identify the local maxima (or minima)
in the CMB maps.
The power spectra of those fields per logarithmic interval in $l$ are then
$l^4 C_l/(2\pi)$ and $l^6 C_l/(2\pi)$, respectively, while the power
spectrum of the temperature fluctuations filed is 
$l^2C_l/(2\pi)$ whose integral over $l$ space produces $C(\theta)$.
Figure \ref{fig:Clln} shows the
power spectra of $\Delta_{,i}$ and $\Delta_{,ij}$ as a function of 
$1/l$ $(\propto \theta)$, and clearly
explains that they strongly enhance the oscillations of $l^2C_l$.
If recalling the relation of $\theta\approx 2\pi/l$, 
there are indeed correspondences between both oscillations of
$\xipk(\theta)$ and $C_l l^2/(2\pi)$; the depression at $\theta\approx
70'$ corresponding to the scales around the first Doppler peak, a
prominent peak at $\theta \approx 13'$, and a damping tail at
$\theta<10'$ associated with the Silk damping in $C_l$.
Most importantly, these oscillatory features of $\xipk(\theta)$ physically 
mean that pairs of hotspots are discretely distributed with some
characteristic separation angles on the last scattering surface. 
The peak statistics thus produces an additional information
that could be convenient for the study of the weak lensing although it
relys on only a subset of available information (the location of peaks,
peak height and their profile). Furthermore, we expect that the
distribution of hotspots are more robust against the systematic
observational errors.
%because it does not need
%to use values of temperature fluctuations at peak posistions.
The previous works already concluded that
the lensing effect on $C_l$ is small at $l<3000$ (e.g. see Seljak 1996
and references therein),
and we therefore expect that the observed $C_l$ will provide the accurate
prediction of unlensed $\xipk(\theta)$. 
The lensing signatures to $\xipk(\theta)$ are then extracted as non-Gaussian
signatures,  which are differences between the observed (lensed) and the predicted
two-point correlation functions. 

\subsection{Lensed two-point correlation function of hotspots}

%%%%%%%%%%%%%%%%%%%%%%%%%%% Figure 6 %%%%%%%%%%%%%%%%%%%%%%%%%%%%%%%%%%%%%
\begin{figure}[t]
 \begin{center}
     \leavevmode\epsfxsize=12cm \epsfbox{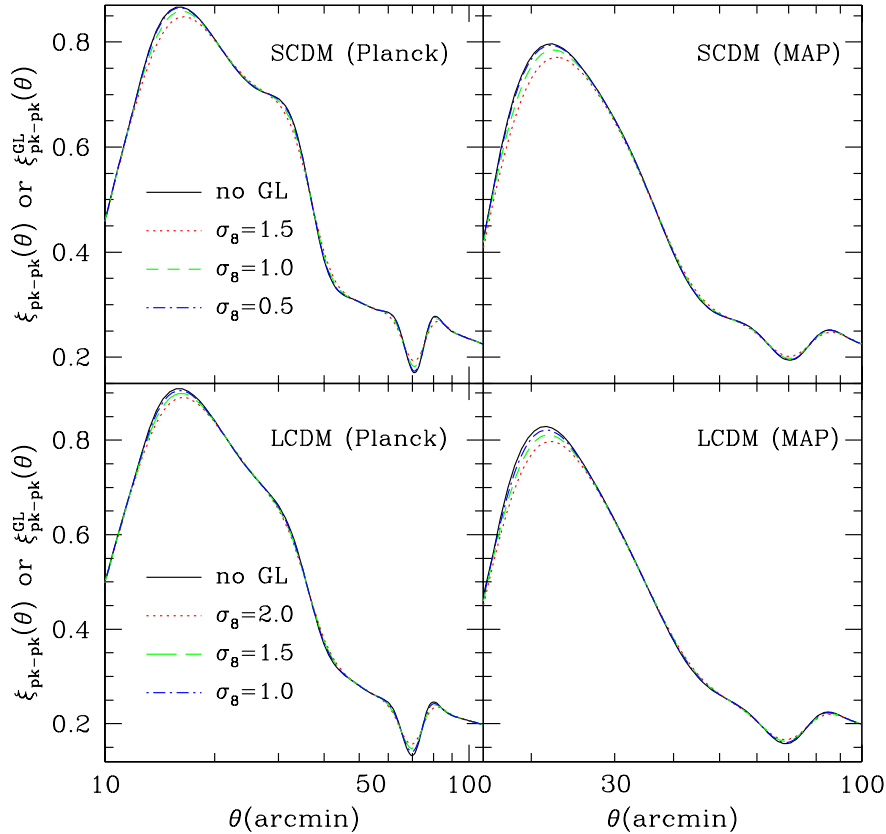}
\caption{
The lensed and unlensed two-point correlation functions of hotspots as a 
function of $\theta$ using
the same parameters of experimental effects as those in Figure
  \ref{fig:xipk}. The weak lensing effects are computed using
  the same models of matter power spectrum with sets of $\sigma_8$ as in 
  Figure \ref{fig:dispgl}. 
 \label{fig:lensxipk} }
 \end{center}
\end{figure}
We present the theoretical predictions of the lensed two-point
correlation function of hotspots. By using equation (\ref{eqn:lensxi}) 
we then compute the lensing effect on the intrinsic $\xipk(\theta)$ 
into which the experimental effects are already included by the
method presented in the previous subsection.
Although the order of computing those secondary effects is not
correct, the treatment makes the calculations much simpler and 
is a good approximation because of
the following reasons. As explained,
the instrumental effects on $\xipk(\theta)$ can be approximately included 
only by modifying the angular power spectrum $C_l$ based on the Gaussian 
theory and actually the predictions are in good agreement 
with the numerical experiments of $\xipk(\theta)$ (HS and also see
Figure \ref{fig:obsxi}).
 This means that the distribution of hotspots as a point process can be 
accurately predicted even in the CMB maps altered by 
the beam smearing and the detector noise.  
Then, since we are here interested 
in a problem how the weak lensing redistributes the 
 `key' hotspots, which can survive
after the beam smearing effect, in an actual observed map,
we can approximately consider the lensing effect on the
$\xipk(\theta)$ after the instrumental effects as long as 
the weak lensing does not create a lot of spurious hotspots on the map. 
Moreover, we can at least say that the important
lensing signature to  $\xipk$  
on large angular scales such as $\theta\approx70'$ is not directly 
coupled to the beam smearing effect of $\theta_{\rm fwhm}\simlt 10'$.
The validity of our treatment has been confirmed by the numerical
experiments on the relevant angular scales. 
%Figure \ref{fig:lensxipk} shows the shapes of lensed $\xipk(\theta)$
%for the same models of
%matter power spectrum as in Figure \ref{fig:dispgl}.
%Figure \ref{fig:lensxipk} shows the weak lensing effect on
%$\xipk(\theta)$ shown in Figure \ref{fig:xipk}. 
Figure \ref{fig:lensxipk} shows both unlensed
and lensed two-point
correlation functions of hotspots, say $\xipk^{\rm GL}(\theta)$ and
$\xipk(\theta)$, where the threshold $\nu=1$
is similarly assumed and we employ the same models of matter power
spectrum as in Figure \ref{fig:dispgl}.
Although $\nu=1$ is assumed throughout this paper
for simplicity,  the two-point correlation function
of hotspots above height of an arbitrary threshold
can be predicted under the Gaussian theory and we could use
this freedom for measuring the lensing effect as will be discussed
later. 
% The experimental
%effects are also taken into account as in Figure \ref{fig:xipk}.
The upper left and bottom left panels for Planck case clearly
demonstrate that the weak lensing effect fairly smooths out the shape
of unlensed $\xipk(\theta)$ (solid line).
The magnitude of the lensing effect is strongly sensitive to the
amplitude of $\sigma_8$ in each cosmological models and
larger at angular scales where $\xipk(\theta)$ has more oscillatory
features (TKF).
In particular, we stress that the weak lensing causes
a distinct smoothing effect on the depression feature of $\xipk(\theta)$
at scales around $\theta\approx 70'$. We therefore 
expect that the observed depth of
depression relative to the plateau shape at larger or smaller 
scales than the scale can be a robust indicator of the weak lensing
signatures and depends only on the magnitude of $\sigma_{\rm GL}$
if it is large sufficiently to detect. 
%We have then confirmed that the depth of the unlensed
%$\xipk(\theta)$ depends mainly on the magnitude of $\Omega_b h^2$. 
However, since in the MAP cases (upper right and bottom right panels)
the detector noise effect decreases largely the depth of the
intrinsic depression as shown in Figure \ref{fig:xipk}, the 
lensing effect is hidden at the scale for all models. 
%For this reason, we will consider the Planck case hereafter. 
If using the relation between $k$ and $l$ shown in Figure \ref{fig:psdkdz}(a),
the measure of $\sigma_{\rm GL}(\theta)$ from the lensing signature
at $\theta\approx 70'$
can provide a constraint on the amplitude of mass fluctuations with
large wavelength modes such as $\lambda\approx 50h^{-1}{\rm Mpc}$. 
On the other hand, the measurement of $\sigma_{\rm GL}$ at
the prominent peak scale such as $\theta\approx 20'$ is sensitive
to the smaller scale structures such as $\lambda\approx
10h^{-1}{\rm Mpc}$, while the shape of $\xipk$ is also 
sensitive to the experimental effects at such scales.

\section{Detectability of the lensing signatures}

In this section, by using numerical experiments
we quantitatively investigate how accurately
non-Gaussian signatures on $\xipk(\theta)$ induced by the weak lensing
can be  detected with the expected future MAP and Planck surveys.

\subsection{Numerical experiments}
%%%%%%%%%%%%%%%%%%%%%%%%%%% Figure 7 %%%%%%%%%%%%%%%%%%%%%%%%%%%%%%%%%%%%%%
\begin{figure}[t]
 \begin{center}
  \leavevmode\epsfxsize=14cm \epsfbox{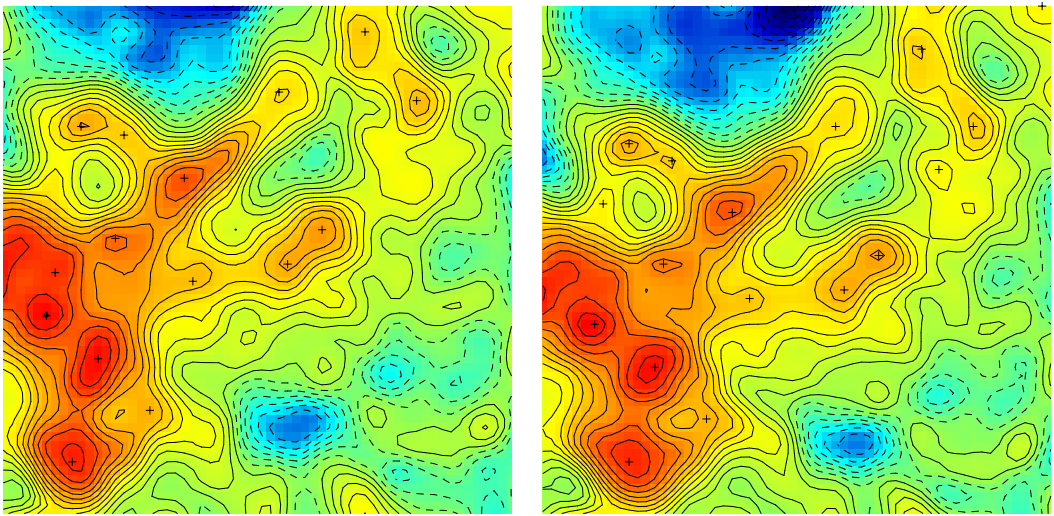}
\caption{
An example of simulations of primordial temperature fluctuations
map (left) and the map deformed by the lensing effect (right) on
a side of $2 $ degree, where we assumed the instrumental
resolutions of Planck
($\theta_{\rm fwhm}=5.5'$ and $\sigma_{\rm sens}=4.3\times10^{-6}$). 
We employed the SCDM model and the lensing displacements
field is computed by using $\sigma_8=1.5$. The peaks above
$1\sigma$ are marked with crosses and the 
contours are stepped in units of $3\sigma_{\rm sens}$. 
\label{fig:cmbmap} }
 \end{center}
\end{figure}
We perform simulations of the CMB maps with and without the lensing effect
by using the following procedures. 
A realistic unlensed temperature maps on a fixed square grid
can be generated from a given power spectrum, $C_l$, based on the
Gaussian assumption. Each simulated map is initially on $40\times40$
square degree area with $4096\times4096$ pixels. After regridding to take
into account the beam smearing effect, the actual pixel number is
reduced to $1200\times1200$ and $480\times480$ for the Planck and
MAP cases, respectively. To compute the
lensing effect on the CMB maps, we first need the convergence field
on each grid. We then
employ the following angular power spectrum of convergence field
(\cite{Blandford}; \cite{miralda}; \cite{Kaiser92}); 
\begin{equation}
 P_\kappa(l)=\frac{9}{4}H_0^4\Omega_{\rm m0}^2
  \int^{\chi_{\rm rec}}_0\!\!d\chi
  a^{-2}W^2(\chi,\chi_{\rm rec})P_{\delta}
  \left(k=\frac{l}{r(\chi)},\chi\right), \label{eqn:pskappa}
\end{equation}
where the convergence field $\kappa$ is expressed in terms of the
radial integral of three dimensional density fluctuations field $\delta$ as
$\kappa =(3/2)H_0^2\Omega_{\rm m0}\int\!d\chi W(\chi,\chi_{\rm rec})
r(\chi)a^{-1}\delta$. 
Note that the second moment of convergence field is then
$\kaco{\kappa^2(\bm{\theta})}=\int(ldl/2\pi)P_\kappa(l)$.
Using the power spectrum $P_\kappa(l)$, the convergence field
can be generated as a realization of a Gaussian process.
To compute the displacement vector $\delta\bm{\theta}$,
we transform the convergence field in the Fourier space, and
compute the Fourier component of the displacement vector,
$\delta\tilde{\bm{\theta}}(\bm{l})$, by using the relation of 
\begin{equation}
\delta\tilde{\bm{\theta}}(\bm{l})=2i
 \frac{\bm{l}}{l^2}\tilde{\kappa}(\bm{l}), 
\end{equation}
where we have used the relation of $2\kappa(\bm{\theta})=\partial(\delta
\theta_1)/\partial \theta_1+\partial(\delta\theta_2)/\partial \theta_2$
derived by the cosmological lens equation. 
If transforming it back to real space, 
for each point on the observed temperature map we can obtain the
corresponding displacement vector to map the point on a irregular grid
of the primary temperature map on the last scattering surface. 
The lensed temperature map can be then obtained by using cloud-in-cell
interpolation to compute the value on the original regular grid of
observed map.
In the case of taking into account the instrumental effects of beam smearing
and noise, we furthermore smooth out the temperature map by convolving
the Gaussian window function ${\cal F}(\bm{\theta},\bm{\theta}')
=1/(2\pi\theta_s^2)\exp[-|\bm{\theta}-\bm{\theta}'|^2/(2\theta_s^2)]$ and
 then add randomly the noise field into each pixel.
Figure \ref{fig:cmbmap} illustrates an example of realizations of
the simulated unlensed and lensed maps for the SCDM model with $\sigma_8=1.5$,
which provides the largest lensing signatures of our considered models. 
Interestingly, the figure shows that the displacement of each position
of hotspots in the lensed map from the unlensed map 
is relatively large even though the global patterns of temperature
fluctuations in both maps are not considerably changed.
This is the consequence of the large scale modes of the deflection
angle, which play an important role to the lensing effect on $\xipk(\theta)$.

\subsection{The signal to noise ratio of the lensing signatures}
%%%%%%%%%%%%%%%%%%%%%%%%% Figure 8 %%%%%%%%%%%%%%%%%%%%%%%%%%%%%%%%%%%%%%%
\begin{figure}[t]
 \begin{center}
     \leavevmode\epsfxsize=14cm \epsfbox{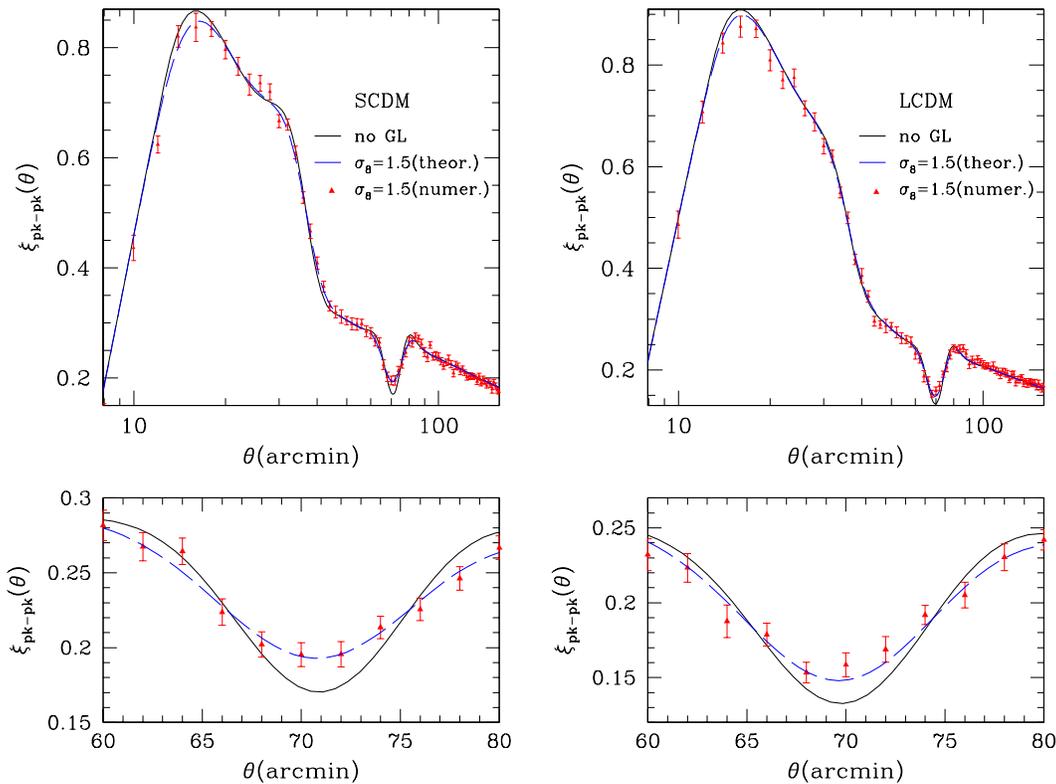}
\caption{
 An example of averaged two-point correlation function with lensing
  effect expected from Planck survey in SCDM (left) and LCDM (right)
  models with $\sigma_8=1.5$, respectively.
  The solid and dashed lines show the
  theoretical predictions of unlensed and lensed $\xipk(\theta)$ as in
  Figure \ref{fig:xipk} and \ref{fig:lensxipk}, respectively.
  The numerical results are obtained from one set of 17 realizations of
  the lensed maps, and the error bars correspond to the cosmic variance
  computed from $100$ realizations. The bottom panels show the
  results around the depression angular scale ($\approx 70'$)
  of the unlensed $\xipk(\theta)$. 
\label{fig:obsxi}
}
 \end{center}
 \end{figure}
The observational errors associated with measurements of $\xipk(\theta)$
arise from the cosmic variance and the instrumental resolutions.
To accurately compute the cosmic
variance, we have used $100$ independent realizations of both
unlensed and lensed
temperature maps, respectively, with $40\times 40$ square degree area.
In this paper we assume $65\%$ sky coverage for MAP and Planck
surveys, and this corresponds to the assumption that 
we can use about $17$ independent simulated maps for those surveys 
in order to obtain the averaged $\xipk(\theta)$. 
In Figure \ref{fig:obsxi} we show an example of
the averaged {\em lensed} two-point
correlation functions of hotspots computed from one set of
17 realizations for the 
expected Planck runs, where the error-bars in each bin can be estimated
by rescaling the variance of the estimates obtained from the $100$
realizations by a factor of $\sqrt{17}$ and the resolution of bins 
in $\theta$ is limited by the pixel size of simulated CMB maps.
The unlensed cases have been already presented by HS.
The bottom panels also show the results around the depression
scale $\theta\approx70'$ of $\xipk(\theta)$ in each cosmological model.
Figure \ref{fig:obsxi} clearly shows that the theoretical predictions
are in remarkable good agreement with the numerical results. 
In particular, the lensing signatures at the depression scale
definitely deviate from the unlensed cases, and  we therefore expect
that the non-Gaussian signatures should be detected by
Planck with high significance for some adequate values of $\sigma_8$.
On the other hand, for the lensing signatures at smaller scales such as
the prominent scale $\theta\approx20'$, it seems to be slightly
difficult to distinguish them because the sampling variance are
larger at smaller scales 
and the shapes of $\xipk(\theta)$ is also sensitive to the instrumental
effects of beam smearing and noise as shown in Figure \ref{fig:xipk}. 
On the other hand, we concluded that it is
difficult for MAP survey to distinguish
the lensing signatures on $\xipk(\theta)$ for all cosmological models
considered in this paper. 
%because of the large beam size and noise level as shown by Figure
%\ref{fig:lensxipk}. 
For this reason, we present only results from Planck in the following.
%However, Figure \ref{fig:obsxi}
%indicates  taht the errors are larger at smaller scales because of the
%smapling variance, and thus
%it seems to be difficult to distinguish the lensing signatures at the small
%scales such as the prominent peak scale $\theta\approx 20'$. 

We so far have adopted the specific shape of matter power spectrum 
in each considered cosmological model,  and therefore
a free parameter of our model is only $\sigma_8$.
%However, the actual observable quantity of our method is the lensing
%dispresion, and therefore we have to carefully explore how
%cosmological information can be extracted from the dispersion.
While the actual observable quantity of our
method is the lensing dispersion $\sigma_{\rm GL}(\theta)$,
we here investigate the dependence
of the lensing signatures to $\xipk(\theta)$ on $\sigma_8$
for simplicity and the problem will be discussed in the
final section.  In Table \ref{tab:1} we summarize the results obtained for the
determination of $\sigma_8$ with a best fit and
the error associated with this determination.
The error then has been determined as the variance of the best fit values of
$\sigma_8$ among a lot of sets of the numerical experiments performed by
fitting the simulated results to the theoretical templates of lensed 
$\xipk(\theta)$ so that the $\chi$-square is minimum. 
Each best fit of theoretical curves to 
the numerical experiments is restricted mainly from the data at
$\theta>20'$, in particular the data around the depression
scale of $\xipk(\theta)$. 
One can see that the signal to noise ratio indeed grows with $\sigma_8$
in each cosmological model. Furthermore, if comparing results obtained
from SCDM and LCDM models with the same value of $\sigma_8$
($\sigma_8=1.5$), it is clear that
more robust constraints can be provided in SCDM, more generally
for background models with larger $\Omega_{\rm m0}$,  as expected. 
The noise level of Planck is independent
on the $\sigma_8$ estimations, but we have confirmed that
the beam size is rather important for the detections.
Importantly, the accuracies of $\sigma_8$ determinations expected from
Planck reach about $8\%$ and $11\%$ for
models with $\sigma_8=1.5$ in SCDM and LCDM, respectively. 
Furthermore, in the Gaussian random theory the two-point correlation
function of local minima ({\em coldspots})
should have the same shape as that of hotspots, and therefore combing the
measurements of coldspots correlation function improves the lensing
signals by about factor $\sqrt{2}$. We could also improve the
significance by combining the measurements of $\xipk$ with another
different thresholds, but the independence of data
then has to be carefully investigated because the
angular positions of same hotspots are used many times in the fitting.  

%############################ Table Captions ############################
%########################################################################
%############################# Table 1 ##################################
\begin{table}
\centering
 \caption{Summary of best fit of $\sigma_8$ determinations from
 the lensing signatures to $\xipk(\theta)$ using the
 numerical experiments for the expected Planck survey in SCDM and LCDM
 models. We have assumed $65\%$ sky coverage,
and the $1\sigma$  error in each determination
represents uncertanities caused by
 the cosmic variance and the instrumental effects due to the beam
 smearing effect and the detector noise (see text).}
\begin{tabular}{c|cc}
    \hline\hline
   input values of $\sigma_8$ & SCDM $(\Omega_{\rm m0}=1,h=0.5)$ & LCDM
   $(\Omega_{\rm m0}=0.3, \Omega_{\rm \lambda 0}=0.7, h=0.7)$\\
    \hline
    0          &$0.30\pm0.31$&$0.23\pm 0.35$\\
    0.5        &$0.47\pm0.24$& - \\
    1.0        &$1.03\pm0.13$&$0.93\pm0.23$\\
    1.5        &$1.52\pm0.12$&$1.49\pm0.17$\\
    1.5 (without noise)  &$1.47\pm0.12$&$1.53\pm0.16$\\
   2.0         &-&$1.97\pm0.16$\\
   \hline
 \end{tabular}
\label{tab:1}
\end{table}
%#############################################################################

\begin{figure}[t]
\begin{center}
     \leavevmode\epsfxsize=10cm \epsfbox{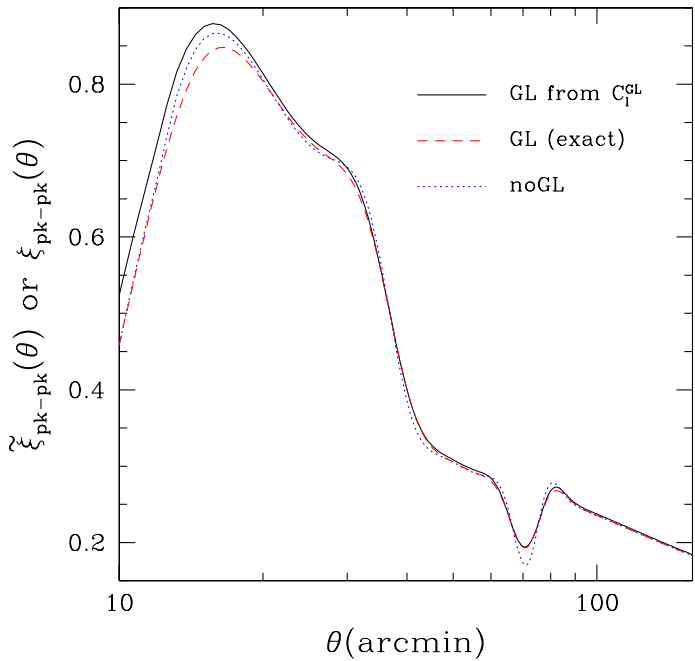}
\caption{This figure shows
the {\em fake} theoretical predction of lensed hotspots correlation
function (solid line), $\tilde{\xi}_{\rm
pk-pk}$, computed from the {\em lensed} (measured) $C_l$ based on
the Gaussian theory in the SCDM model with $\sigma_8=1.5$.
The exact prediction (dashed line) and the unlnesed case (dotted line)
are also shown.
Since the lensed temprature fluctuations field weakly deviates
from the Gaussian, $\tilde{\xi}_{\rm pk-pk}$ can no longer
accurately describe the distribution of hotspots in the lensed maps. 
\label{fig:fakexi}
}
\end{center}
\end{figure}

Our arguments presented in this section
rely on the expectation that we can accurately
predict the shape of unlensed $\xipk(\theta)$ as a function of sets
of cosmological parameters constrained from the measured angular
power spectrum $C_l$ within the limit of the observational errors.
However, we should bear in mind the fact that only the lensed $C_l$ is
measurable. Then, one may imagine an approach to
compare the measured $\xipk$ to the {\em fake} prediction,
say $\tilde{\xi}_{\rm pk-pk}$, computed directly
from the lensed (measured) $C_l$ based on the Gaussian theory. 
Since the lensed temperature fluctuations field
weakly deviates from the Gaussian (\cite{Bern97}), the
distribution of hotspots in the lensed maps can be no longer
characterized only by the lensed $C_l$.
However, it will be still interesting to see differences between
the exact and fake predictions of the lensing effect on
$\xipk(\theta)$.
Figure \ref{fig:fakexi} thus shows the shape of $\tilde{\xi}_{\rm
pk-pk}$ and  reveals that $\tilde{\xi}_{\rm pk-pk}$
overestimates the power of correlations on scales of $\theta<40'$,
while $\tilde{\xi}_{\rm pk-pk}$
mimics the smoothing effect on the depression feature
of $\xipk$ resulting from the fact that the weak lnsing
already causes the smoothing of $C_l$.  
Consequently, for example, the value of reduced $\chi$-square
for the fitting between $\tilde{\xi}_{\rm pk-pk}$ and
the simulated $\xipk$ becomes worse to be $\chi^2\approx 162/95$
against the
fact that the fitting of using the exact theoretical
predictions (\ref{eqn:lensxi}) of lensed $\xipk$ produces
values around almost unity.
%This means that the lensed $\xipk$ that can be
%directly measured in the lensed maps
%cannot be precisely predicted only from the measred $C_l$ based on the
%Gaussian theory.
Hence, the important problem how we extract
the lensing information only from the measurable CMB quantities
has to be carefully investigated and
this problem will be again discussed in the next section. 

%########################## Figure 9 ###################################

\section{Discussion and Conclusions}
\label{summary}
In this paper, we have quantitatively shown that the peak-peak
correlation function for the CMB maps can be an efficient measure to
probe the mass fluctuations at relatively large scales.
The non-Gaussian signatures on $\xipk(\theta)$ caused by the weak
lensing appear at several scales as smoothing effects on
the oscillatory features of $\xipk(\theta)$ (TKF). In particular, 
by using numerical experiments of the CMB maps including effects of
beam smoothing and detector noise as well as the lensing displacements,
we found that the lensing signatures at about {\em one degree} scales,
where the intrinsic $\xipk(\theta)$ has the pronounced depression
feature, should be detected most significantly from the expected Planck
survey if the signatures are adequately large for the detections.
On the other hand, unfortunately we concluded that
it is difficult for MAP to detect the lensing
signatures for the cosmological models considered in this paper.
The direct observable quantity of our method is the dispersion of
lensing deflection angle as shown by equation (\ref{eqn:lensxi}).
We have then revealed that the one-degree
scale dispersions are sensitive to amplitudes of three dimensional
mass fluctuations around
wavelength $\lambda\approx 50h^{-1}{\rm Mpc}$ because of the
projection effect. Furthermore, such lensing dispersions have
contributions from structures of the universe with a wide redshift
distribution in the range of $1\simlt z\simlt 3$ (see Figure
\ref{fig:psdkdz}). This projection effect would be thus a serious issue
for extracting the cosmological information from weak lensing
contributions in a general case. However, since the mass fluctuations
around $\lambda\sim 50h^{-1}{\rm Mpc}$ are now still in the linear
regime and the evolution history in the redshift space
is theoretically well understood in the context of
gravitational instability in an expanding universe, the lensing
signatures can accurately determine the cosmological
parameters associated with amplitude and evolution of 
mass fluctuations at the scales.  It is therefore expected that
our method can provide robust constraints on cosmological parameters
$\sigma_8$ and $\Omega_{\rm m0}$ without much specifying the shape 
of matter power spectrum. Our numerical
experiments indeed revealed that significant signal to noise
rations for determinations of $\sigma_8$ from the lensing signatures to
$\xipk(\theta)$ are obtained for some input values of $\sigma_8$
in SCDM and LCDM models, respectively. It was also shown that for the same
value of $\sigma_8$ more significant signal can be obtained in SCDM than 
in LCDM.  To break the degeneracy in $\sigma_8-\Omega_{\rm m0}$
determinations only by using our method,
one must measure the scale dependence of
lensing signatures. This seems to be difficult even for Planck survey,
because the shape of $\xipk(\theta)$
at small scales of $\theta\approx20'$ where $\xipk$ has the secondly
significant lensing signatures
%the small scale lensing effects on $\xipk(\theta)$ required 
%for the breaking
is also sensitive to the instrumental effects of
beam smoothing and noise.
%Therefore, we have to invoke the other methods for the breaking.
Anyway, it is very interesting that our method could provide constraints on
the mass fluctuations in the linear regime independently
of those provided by the survey of galaxies clustering and the primary
CMB anisotropies alone, which cannot directly probe the power spectrum of
dark matter.

In the results presented in this paper, we have discussed how accurately
the non-Gaussian signatures of the lensed $\xipk^{\rm GL}$ can be
detected as a deviation from the unlensed $\xipk$ as shown in
Figure \ref{fig:obsxi}. This strategy is based on the
assumption that we can accurately predict the
unlensed $\xipk(\theta)$ from the measured $C_l$
because the lensing effect on $C_l$ is small (\cite{Seljak96}).
However, since an actual measurable quantity is only the lensed $C_l$,
there remains an important problem that we have to carefully
investigate. Usually,
the measured angular power spectrum $C_l$ can be used to constrain
the sets of cosmological parameters under a specific scenario,
for example, within the framework of inflationary-motivated models
(e.g. \cite{lange}). 
The detailed analysis of using higher $l$ modes such as $l>3000$
will need to take into account the lensing contributions to $C_l$ for
the accurate determinations (\cite{Stompor}).
Therefore, one approach toward detecting
 the lensing effect on $\xipk$ is
that we first construct a lot of templates of the unlensed $\xipk$
as a function of sets of cosmological parameters constrained by
the measured $C_l$ within the observational errors,
and then we compare the templates with the directly measured
$\xipk$ in the observed (lensed) CMB maps by taking into account the lensing
contributions. Based on the considerations, we are now investigating the 
detailed dependence of cosmological parameters on the shape of
$\xipk$. As a result, we confirm to a extent that, if we focus on the
depth of depression feature of $\xipk(\theta)$ at one degree scales
relative to the plateau shapes at larger or smaller scales than that
scale, the measured depth could be a robust indicator relatively
independently of the cosmological parameters because
the weak lensing with same magnitude of $\sigma_{\rm GL}$
can shallow the measured depth to a same amount.
%to same extent.  
%the measured depth is shallowed to same extent for 
%the lensing dispersions $\sigma_{\rm GL}$ with the same magnitude.
This should be further investigated carefully and will be
presented elsewhere. 
The important thing that we stress in this paper
is that we quantitatively proposed a new statistical
method for measuring the lensing effect based on the peak statistics.
It is not evident that our method and
the other methods can measure
the small
lensing signatures from the observed CMB data with exactly {\em same}
statistical significance. Therefore, several independent methods
should be performed to measure the lensing signatures
complementarily.

Undoubtedly, we have to carefully consider how secondary anisotropies
and the foreground sources affect our conclusions. The most important
source of secondary anisotropies is the (thermal) Sunyaev-Zel'dovich (SZ)
effect, which is essentially caused by hot electrons inside 
the clusters of galaxies. Since SZ induces redistributions of the
photon energy from low frequency to high frequency part of the black
body spectrum, the effect could mimic peaks in the observed temperature
map. However, the SZ anisotropies are dominated by the Poisson
contribution from the individual clusters at relevant angular
scales (\cite{KK}), and therefore
the contribution of the cross correlation between spurious
peaks due to SZ and the intrinsic peaks will be smaller
than the amplitude of the intrinsic peak-peak correlation because
the peaks on the last scattering surface and the SZ clusters are
statistically uncorrelated. 
Furthermore, we should emphasize that the SZ effect can be always removed 
by either observing at $217$ GHz or by taking advantage of its
specific spectral properties. The other secondary anisotropies such
as Rees-Sciama effect do not affect our method  because the
amplitudes of those effects have a very small contribution
at $l<3000$ (\cite{SeljakRees}).
Finally, with regard to the effect of the extragalactic sources, 
since it is expected that the amplitude of anisotropies due to
the discrete sources in the $100-200$ GHz range are well bellow the
amplitude of primordial fluctuations (\cite{Toffolatti})
and the sources will be also eliminated by using the multi frequency
observations in principle, it can be safely concluded that 
this is not a serious for our method. 

Recent measurements of cosmic shear (e.g. see
\cite{Ludovic}) have provided constraints on the combination
 of $\sigma_8$ and $\Omega_{\rm m0}$. The
cosmic shear can probe the mass fluctuations at
angular scales of $1'\simlt \theta \simlt 30'$ because of the limited
survey volume, where the nonlinear
clustering effect of dark matter could play an important role.
It has been also shown that the deformation effects on the CMB maps
could be measured by Planck (\cite{Bern98}; \cite{WBB}),
and it is sensitive to the projected mass fluctuations
at scales around $5'$ which is the characteristic curvature scale of
the primary temperature field. Therefore, 
although $\Omega_{\rm m0}$ parameter cannot be determined
accurately form the primary CMB anisotropies alone even with
Planck because of the so-called cosmic (geometrical) degeneracy
(\cite{BET}),  it is expected that our method and
those other independent methods of using the weak lensing will break
the cosmic degeneracy with high precision
because the amplitude of lensing contributions
is also sensitive to $\Omega_{m0}$ as explained. Another challenging
possibility is that those independent methods could
allow us to observationally 
reconstruct the shape of power spectrum of dark matter including the
evolution history in the redshift space by
combining those measurements of amplitudes of mass fluctuations at respective
angular scales.

\section*{Acknowledgments}
We thank E. Komatsu for careful reading of the manuscript,
frequent discussions and critical comments.
We also thank an anonymous referee for useful comments which
have considerably improved this manuscript. 
We are also grateful to U. Seljak and M. Zaldarriaga 
for their available CMBFAST code. M.T. thanks to the
Japan Society for Promotion of Science (JSPS) Research Fellowships
for Young Scientist. 

\begin{appendix}
\section{Mean density of hotspots in the 2D CMB map}
\label{app:num}

In this appendix, we briefly review the derivation of the mean
number density of peaks in the two dimensional Gaussian field
(\cite{BBKS,BE}).

Following BBKS and BE, we introduce the spectral parameters defined by
\begin{equation}
\gamma\equiv\frac{\sigma_1^2}{\sigma_0\sigma_2},\hspace{4em}
 \theta_\ast=\sqrt{2}\frac{\sigma_1}{\sigma_2},
\end{equation}
where
\begin{equation}
\sigma_n^2\equiv\int\!\!\frac{ldl}{2\pi}C_l l^{2n}.
\end{equation}
The characteristic curvature scale of primary temperature field can be
estimated as $\theta_\ast\sim5'$ for the cosmological models considered
in this paper.

The problem of the two dimensional peak statistics is to consider the
statistical properties of the point process. We can then express the
point process entirely in terms of the field and its derivatives. 
At hotspots (coldspots) the gradient $\Delta_i(\equiv\partial
\Delta/\partial \theta^i)$ vanishes, and the
eigenvalues of curvature matrix $\Delta_{ij}(\equiv
\partial^2\Delta/(\partial \theta^i\partial\theta^j))$ are all negative
(positive). We therefore need to consider six independent
variables $\bm{v}=(\Delta,\Delta_x,\Delta_y,\Delta_{xx},\Delta_{yy}
,\Delta_{xy})$ to specify {\em one} local maximum.
For the Gaussian field, the probability density function (PDF) for
$\bm{v}$ is
\begin{equation}
p_1(\bm{v})=\frac{1}{(2\pi)^3|{\rm det}(M_{ij})|^{1/2}}\exp\left(
 -\frac{1}{2}v_iM^{-1}_{ij}v_j\right),
\end{equation}
where the covariance matrix is $M_{ij}=\kaco{v_iv_j}$ because of
$\kaco{v_i}=0$ in the present case, and $M^{-1}_{ij}$ 
is the inverse matrix of $M_{ij}$. Following HS,
 we introduce the notations for convenience as 
\begin{eqnarray}
&&\nu\equiv \frac{\Delta}{\sigma_0},\hspace{2em}\eta_i\equiv\frac{\Delta_i}
 {\sigma_1} \ (i=x,y)
 \nonumber \\
&& X\equiv-\frac{\Delta_{xx}
 +\Delta_{yy}}{\sigma_2},\hspace{2em}Y\equiv\frac{\Delta_{xx}-\Delta_{yy}}
 {\sigma_2},\hspace{2em}Z\equiv\frac{2\Delta_{xy}}{\sigma_2}.
\label{eqn:gaussvari}
\end{eqnarray}
The non-zero second moments of these variables are
\begin{eqnarray}
&&\kaco{\nu^2}=\skaco{X^2}=2\skaco{Y^2}=2\skaco{Z^2}
=2\kaco{\eta_i^2}=1,\hspace{1em} \skaco{\nu X}=\gamma,
\end{eqnarray}
where $\gamma\equiv \sigma_1^2/(\sigma_0\sigma_2)$. 
By using these equations, we can obtain the following PDF for the variables
$\bm{v}=(\nu,X,Y,Z,\eta_i)$ with the simple form, and it
gives a probability that the field point has values in the ranges of
$\nu$ to $\nu+d\nu$, $X$ to $X$ to $X+dX$ and so on:
\begin{eqnarray}
p_1(\nu,X,Y,Z,\bm{\eta}_i)d\nu dXdYdZd^2\eta_i
&=&\frac{2^2}{(2\pi)^3\sqrt{1-\gamma^2}}\exp\left[-Q\right]
d\nu dXdYdZd^2\eta_i
\label{eqn:1ppdf}
\end{eqnarray}
with
\begin{equation}
2Q=\frac{(\nu-\gamma X)^2}{1-\gamma^2}+X^2+2Y^2+2Z^2+2\bm{\eta}^2.
\end{equation}

The point process of hotspots (or coldspots) can be described by the
number density `operator'
\begin{equation}
n_{\rm pk}(\bm{\theta})=\sum_p\delta^{D}(\bm{\theta}-\bm{\theta}_{{\rm pk},p}),
\label{eqn:sumpk}
\end{equation}
where $\delta^{D}(x)$ is the Dirac delta function. In the neighborhood
of a hotspot point $\bm{\theta}_{\rm pk}$ we can expand the
field $\Delta(\bm{\theta})$
in a Taylor series:
\begin{equation}
\Delta(\bm{\theta})\approx\Delta(\bm{\theta}_{\rm pk})+\frac{1}{2}\Delta_{ij}
 (\bm{\theta}_{\rm pk})
 (\theta-\theta_{\rm pk})_i(\theta-\theta_{\rm pk})_j. 
\end{equation}
Using this equation, therefore, the number density field can be
expressed in terms of the Gaussian variables:
\begin{equation}
 n_{\rm pk}(\bm{\theta})=|{\rm det}(\Delta_{ij}(\bm{\theta}))|
  \delta^D(\Delta_i(\bm{\theta})). \label{eqn:pkgauss}
\end{equation}
The summation symbol in equation (\ref{eqn:sumpk}) can be eliminated
because the delta function of the above equation
picks out all of the extremal
points which are zero of $\Delta_i(\bm{\theta})$ are
maximum in the two dimensional map. 

Hence the ensemble average of equation (\ref{eqn:pkgauss}) produces the
differential mean number density of hotspots of height
in the range of $\nu$ and $\nu+d\nu$:
\begin{eqnarray}
\bar{n}_{\rm pk}(\nu)&=&\kaco{|{\rm det}(\Delta_{ij})|\delta^D(\Delta_i)}
=\frac{1}{2\theta_\ast^2}\kaco{|X^2-Y^2-Z^2|\delta^D(\eta_i)} \nonumber \\
 &=&\frac{1}{2\theta_\ast^2}\int^\infty_0\!\!dX\int^X_{-X}\!\!dY
  \int^{\sqrt{X^2-Y^2}}_{-\sqrt{X^2-Y^2}}dZ(X^2-Y^2-Z^2)
  p_1(\nu,X,Y,Z,\eta_i=0),
\end{eqnarray}
where we have adopted the conditions for a hotspot of $X>0$ and
$Y^2+Z^2<X^2$. BE derived the analytical expression for
$\bar{n}_{\rm pk}(\nu)$; 
\begin{equation}
\bar{n}_{\rm pk}(\nu)=\frac{1}{(2\pi)^{3/2}\theta^2_\ast}
 \exp(-\nu^2/2)G(\gamma,\gamma\nu),
 \label{eqn:meannum}
\end{equation}
where
\begin{eqnarray}
G(\gamma,x_\ast)&\equiv&(x_\ast^2-\gamma^2)\left\{1-\frac{1}{2}
					    {\rm erfc}\left[
 \frac{x_\ast}{\sqrt{2(1-\gamma^2)}}\right]\right\}+x_\ast(1-\gamma^2)
\frac{\exp\{-x_\ast^2/
[2(1-\gamma^2)]\}}{\sqrt{2\pi(1-\gamma^2)}}\nonumber \\
&&+\frac{\exp[-x_\ast^2/(3-2\gamma^2)]}{(3-2\gamma^2)^{1/2}}
\left\{1-\frac{1}{2}{\rm erfc}\left[\frac{x_\ast}{\sqrt{2
(1-\gamma^2)(3-2\gamma^2)}}\right]
 \right\}.
 \label{eqn:diffnum}
\end{eqnarray}
In this paper, we have often used the mean number density of peak of
height above a certain threshold $\nu_{\rm thresh}$ obtained
by integrating equation (\ref{eqn:meannum}) over $\nu>\nu_{\rm thresh}$. 
Naturally, the mean number density of coldspots
of height below $-\nu$ ($\Delta_{\rm pk}<-\nu\sigma_0$)
is symmetrically
given by $\bar{n}_{\rm cold}(<-\nu)=\bar{n}_{\rm pk}(>\nu)$.

\section{The unlensed two-point correlation function of hotspots}
\label{app:unlensxi}

Based on the peak statistics for the two dimensional Gaussian field,
we briefly review the derivation of two-point correlation function of
hotspots following HS. For this purpose, let us consider
two hotspots separated by
the angular scale $\theta$. In the same way as in appendix \ref{app:num},
we then need the joint probability density
function for the 12 independent variables $\bm{v}=(\bm{v}_1,\bm{v}_2)
=(\bm{v}(\bm{\theta}_1),\bm{v}(\bm{\theta}_2))$
with $\theta=|\bm{\theta}_1-\bm{\theta}_2|$. If using the variables
(\ref{eqn:gaussvari}) for each hotspot,
we can block the covariance matrix $M_{ij}$ for the 12 variables in the order
$(\nu_1,X_1,\eta_{1x},Y_1,\nu_2,X_2, \eta_{2x},Y_2,\eta_{1y}
,\eta_{2y},Z_1,
Z_2 )$ into the $8\times 8$-matrix $M_{(8)}$ (upper left)
and the $4\times4$-matrix $M_{(4)}$ (bottom right); 
\begin{equation}
M_{(8)}=\left(
\begin{array}{cccccccc}
1&\tilde{\gamma}&0&0
 &\lambda_{000}& \lambda_{020}&-\lambda_{011}&\lambda_{022}\\
\ddots &1&0&0&\lambda_{020}&\lambda_{220}&-\lambda_{121} & \lambda_{222}\\
\ddots &\ddots &1/2&0&\lambda_{011}&\lambda_{121}
  &\frac{1}{2}(\lambda_{110} - \lambda_{112})&\frac{1}{2}(\lambda_{123}
  - \lambda_{121})\\
 \ddots&\ddots&\ddots&1/2& \lambda_{022}&\lambda_{222}&
  -\frac{1}{2}(\lambda_{123} - \lambda_{121})&\frac{1}{2}(\lambda_{220}
  + \lambda_{224})\\
\ddots &\ddots&\ddots&\ddots&1&\gamma&0&0\\
\ddots &\ddots&\ddots&\ddots&\ddots&1&0&0\\
\ddots &\ddots&\ddots&\ddots&\ddots&\ddots&1/2&0\\
\ddots &\ddots&\ddots&\ddots&\ddots&\ddots&\ddots&1/2
\end{array}
    \right),
\end{equation}
\begin{equation}
M_{(4)}=\left(
\begin{array}{cccc}
 1/2 &\frac{1}{2}(\lambda_{110} + \lambda_{112})& 0 &
  -\frac{1}{2}(\lambda_{121} + \lambda_{123})\\
\ddots&1/2&-\frac{1}{2}(\lambda_{121} + \lambda_{123})&0\\
\ddots &\ddots&1/2&\frac{1}{2}(\lambda_{220} - \lambda_{224})\\
\ddots &\ddots&\ddots&1/2
\end{array}
    \right),
\end{equation}
where we have introduced the notations (HS):
\begin{eqnarray}
&&\lambda_{ijn}\equiv\frac{1}{\sigma_i\sigma_j}\int
 \frac{ldl}{2\pi}l^{i+j}C_lJ_n(l\theta). \label{eqn:cov2}
\end{eqnarray}
Note that both $M_8$ and $M_4$ are symmetric.

The joint PDF for the two hotspots can thus be expressed in terms of
the above covariant matrix;
\begin{equation}
p_2(\bm{v}_1,\bm{v}_2)=\frac{1}{(2\pi)^6|{\rm det}(M_{ij})|^{1/2}}
 \exp\left(-\frac{1}{2}v_iM^{-1}_{ij}v_j\right). 
\end{equation}
Therefore, as in the similar way of the derivation of mean number density,
we can calculate the unlensed two-point correlation function of hotspots 
which are separated by $\theta$ and have heights above $\nu_1$ and
$\nu_2$, respectively;
\begin{eqnarray}
1+\xipk(\theta)&=&\frac{1}{\bar{n}_{\rm pk}(>\nu_1)\bar{n}_{\rm pk}(>\nu_2)}
\kaco{n_{\rm pk}(\bm{\theta}_1)n_{\rm pk}(\bm{\theta}_2)}_{|\bm{\theta}_1
-\bm{\theta}_2|=\theta} \nonumber \\
&=&\frac{1}{4\theta_\ast^4
 \bar{n}_{\rm pk}(>\nu_1)\bar{n}_{\rm pk}(>n_2)}
 \kaco{(X^{\prime 2}_1-Y^{\prime 2}_1-Z^{\prime 2}_1)
 (X^{\prime 2}_2-Y^{\prime 2}_2-Z^{\prime 2}_2)\delta^D(\eta'_{1i})
 \delta^D(\eta'_{2i})}_{|\bm{\theta}_1-\bm{\theta}_2|=\theta}
\nonumber \\
 &=&\frac{1}{4\theta_\ast^4
 \bar{n}_{\rm pk}(>\nu_1)\bar{n}_{\rm pk}(>n_2)}
 \int_{\nu_1}^\infty\!\!d\nu_1'\int_{\nu_2}^\infty\!\!d\nu_2'
 \int_0^\infty\!\!dX_1'\int_0^\infty\!\!dX_2'\int^{X_1'}_{-X_1'}\!\!dY_1'
 \int^{X_2'}_{-X_2'}\!\!dY_2' \nonumber \\
 &&\times\int^{\sqrt{X_1^{\prime 2}-Y_1^{\prime 2}}}
  _{-\sqrt{X_1^{\prime 2}-Y_1^{\prime 2}}}\!\!dZ_1'
  \int^{\sqrt{X_2^{\prime 2}-Y_2^{\prime 2}}}
  _{-\sqrt{X_2^{\prime 2}-Y_2^{\prime 2}}}\!\!dZ_2'
  (X^{\prime 2}_1-Y^{\prime 2}_1-Z^{\prime 2}_1)
 (X^{\prime 2}_2-Y^{\prime 2}_2-Z^{\prime 2}_2)\nonumber \\
&& \times p_2(\nu_1',X_1',Y_1',Z_1',\eta'_{1i}=0,\nu_2',X_2',Y_2',Z_2',
 \eta'_{2i}=0). \label{eqn:unlensxi}
\end{eqnarray}
To obtain $\xipk(\theta)$, we performed a six dimensional numerical
integration of the equation (\ref{eqn:unlensxi}) because the two
integrals can be done analytically.

\section{Alternative representation of the lensed hotspots correlation
 function}
\label{app:xianaly}

The lensing dispersion $\sigma_{\rm GL}(\theta)$ generally has isotropic 
and anisotropic contributions, which are expressed by
$\sigma_{\rm GL,0}(\theta)$ and
$\sigma_{\rm GL,2}(\theta)$, respectively, as shown in equation
(\ref{eqn:gldisp}). For the standard power spectra of mass
fluctuation as adopted in this paper, the contribution of
$\sigma_{{\rm GL},2}(\theta)$ is smaller than that of
$\sigma_{{\rm GL},0}(\theta)$. If the neglecting the contribution,
the key equation (\ref{eqn:lensxi}) of this paper can be further
simplified. In this case, the equation (\ref{eqn:lensxi}) becomes
\begin{eqnarray}
\tilde{\xi}_{\rm pk-pk}(\theta)\approx
 . \label{eqn:xianaly}\int\!\!d\theta'\theta'\int\!\!ldl
 \xi_{\rm pk-pk}(\theta')
 J_{0}(l\theta')\exp\left[-\frac{l^2}{2}\sigma^2_{\rm GL,0}(\theta)\right]
 J_0(l\theta), 
\end{eqnarray}
and this can be then rewritten as
\begin{equation}
\tilde{\xi}_{\rm pk-pk}(\theta)=\int\!\!\theta'd\theta'\xi_{\rm pk-pk}(\theta')
 K(\theta,\theta'),
\end{equation}
where the kernel $K(\theta,\theta')$ is given by
\begin{equation}
 K(\theta,\theta')\equiv \int^{\infty}_0\!\!ldl J_0(l\theta)J_0(l\theta')
  \exp\left[-\frac{l^2}{2}\sigma^2_{{\rm GL},0}(\theta)\right]
=\frac{1}{\sigma^2_{{\rm GL},0}(\theta)}
\exp\left[-\frac{\theta^2+\theta'^{2}}{2\sigma^2_{{\rm
GL},0}(\theta)}\right]I_0\left[\frac{\theta\theta'}{\sigma^2_{{\rm
GL},0}(\theta)}\right].
\end{equation}
$I_0(x)$ is the modified zeroth-order Bessel function, and we have 
used equation (6.633.2) of Gradshteyn \& Ryzhik (1994).
As shown in Figure \ref{fig:dispgl}, $\sigma_{\rm GL,0}(\theta)/\theta<1$ and
the argument of $I_0$ is generally a large number. Therefore, if using
the approximation $I_0(x)\approx (2\pi x)^{-1/2}\exp(x)$ for
$x\rightarrow \infty$, we can rewrite equation (\ref{eqn:xianaly}) in
the following form;
\begin{equation}
\xi^{\rm GL}_{\rm pk-pk}(\theta)\approx
 \frac{1}{(2\pi \theta)^{1/2}\sigma_{\rm GL,0}(\theta)}
 \int\!\!\theta'd\theta'\xi_{\rm pk-pk}(\theta')\exp
 \left[-\frac{(\theta-\theta')^2}
	{2\sigma_{\rm GL,0}^2(\theta)}\right]
 \end{equation}
This expression explicitly explains that
the lensing effect on $\xipk(\theta)$ appears as a Gaussian smoothing
with relative width $\sigma_{\rm GL,0}$, and then the asymptotic behavior at
$\sigma_{\rm GL,0}\rightarrow 0$ is naturally  $\xipk^{\rm GL}
(\theta)\rightarrow \xipk(\theta)$. 
\end{appendix}

\clearpage
%#############################################################################
%########################## Figure captions ##################################
%##############################################################################

\end{document}